%% file: rqe_main.tex
\documentclass{article}

\usepackage{arxiv}

\usepackage[utf8]{inputenc} 
\usepackage[T1]{fontenc}    
\usepackage{booktabs}       
\usepackage{amsfonts}       
\usepackage{nicefrac}       
\usepackage{microtype}      
\usepackage{lipsum}
\usepackage{cite}
\usepackage{amsmath,amssymb,amsfonts}
\usepackage{algorithmic}
\usepackage{graphicx}
\usepackage{textcomp}
\usepackage{multirow}
\usepackage{multicol}
\usepackage{xcolor}

\newcommand{\shrt}{Power Swapper}

\newcommand{\comm}[1]{\textcolor{red}{#1}}
\newcommand{\tx}[1]{\textit{#1}}

\usepackage{listings}
\definecolor{codegreen}{rgb}{0,0.6,0}
\definecolor{codegray}{rgb}{0.5,0.5,0.5}
\definecolor{codepurple}{rgb}{0.58,0,0.82}
\definecolor{backcolour}{rgb}{0.95,0.95,0.92}

\lstdefinestyle{mystyle}{
    backgroundcolor=\color{white},   
    commentstyle=\color{codegreen},
    keywordstyle=\color{magenta},
    numberstyle=\tiny\color{codegray},
    stringstyle=\color{codepurple},
    rulecolor=\color{black},
    frame=shadowbox,
    basicstyle=\ttfamily\footnotesize,
    breakatwhitespace=false,         
    breaklines=true,                 
    captionpos=b,                    
    keepspaces=true,                 
    numbers=left,                    
    numbersep=5pt,                  
    showspaces=false,                
    showstringspaces=false,
    showtabs=false,                  
    tabsize=2
}

\def\BibTeX{{\rm B\kern-.05em{\sc i\kern-.025em b}\kern-.08em
    T\kern-.1667em\lower.7ex\hbox{E}\kern-.125emX}}
\begin{document}

\title{Reverse Engineering of Integrated Circuits: Tools and Techniques\\
}

\author{Abhijitt Dhavlle \\
Electrical and Computer Engineering \\
\textit{George Mason University}\\
Fairfax, USA. \\
adhavlle@gmu.edu}

\maketitle

\begin{abstract}
Consumer and defense systems demanded design and manufacturing of electronics with increased performance, compared to their predecessors. As such systems became ubiquitous in a plethora of domains, their application surface increased, thus making them a target for adversaries. Hence, with improved performance the aspect of security demanded even more attention of the designers. The research community is rife with extensive details of attacks that target the confidential design details by exploiting vulnerabilities. The adversary could target the physical design of a semiconductor chip or break a cryptographic algorithm by extracting the secret keys, using attacks that will be discussed in this thesis. This thesis focuses on presenting a brief overview of IC reverse engineering attack and attacks targeting cryptographic systems. Further, the thesis presents my contributions to the defenses for the discussed attacks. 
The globalization of the Integrated Circuit (IC) supply chain has rendered the advantage of low-cost and high performance ICs in the market for the end users. But this has also made the design vulnerable to over production, IP Piracy, reverse engineering attacks and hardware malware during the manufacturing and post manufacturing process. Logic locking schemes have been proposed in the past to overcome the design trust issues but the new state-of-the-art attacks such as SAT has proven a larger threat. This work highlights the reverse engineering attack and a proposed hardened platform along with its framework. On the other side, the side-channel attacks (SCAs) has been one of the emerging threats. These SCAs function by exploiting
the side-channels which invariably leak important data during
an application’s execution. The information leaked through side-channels are 
inherent characteristics of the system and is often unintentional. 
This information can be 
microarchitectural or physical information such as 
power consumption, thermal maps, timing of the operation, acoustics, 
and cache-trace. 
Intercepting secret information based on the study of power signature is a subdivision of SCAs where power consumption information serves as a covert channel leaking crucial information about the executed operations. 
Such physical SCAs are known to be a significant threat to 
cryptosystems such as AES (Advanced Encryption Standard) and can reveal the encryption key efficiently. 
To overcome such concerns and protect the
data integrity, I introduce \textit{\shrt\ } in this work. The
proposed \shrt\ thwarts the attack by randomly choosing one of the multiple modules that perform the intended activity, but have  power signature different than a standard implementation and can lead to similar power consumption as one of the other modules that perform a different operation. 
To achieve this, I introduce carefully crafted swapping of the standby modules that are responsible for the AES operation thus deluding the attacker without hurting the crypto operation. 
This methodology has been validated for the AES power analysis attack and the key information observed by the attacker is seen to
be incorrect, indicating the success of the proposed method.

\end{abstract}


\include{chapterOne}
\include{chapterTwo}
\include{ChapterThree}
\include{conclusion}

\bibliographystyle{IEEEtran}
\bibliography{rqe_main}
\end{document}

%% file: chapterOne.tex
\section[Introduction]{Introduction}
\subsection{Introduction to Hardware Security}
Consumer and defense systems demanded design and manufacturing of electronics with increased performance, compared to their predecessors \cite{Sathwika_DT'22}. As such systems became ubiquitous in a plethora of domains, their application surface increased, thus making them a target for adversaries \cite{SK_1,SK_2,SK_3,SS_1,SS_2,SS_3,SS_4,Dhavlle_date'21,SS_6,SS_7,SS_8,SS_9,SS_10}. Hence, with improved performance the aspect of security demanded even more attention of the designers. Confidentiality, Integrity and Availability are the major building blocks of the state-of-the-art systems. The confidentiality refers to the design confidentiality; and integrity refers to the design integrity where an adversary should not make malicious modifications to the design; whereas,  availability refers to the system functioning as intended by the designer. Hence, to maintain these tripod principles, hardware security, a collection of various defense methodologies, emerged. The research community is rife with extensive details of attacks that target the confidentiality and integrity of the device. The adversary could target the physical design of a semiconductor chip or break a cryptographic algorithm by extracting the secret keys, using attacks that will be mentioned shortly. This thesis focuses on presenting a brief overview of IC counterfeiting attacks and attacks targeting cryptographic systems. Further, the thesis presents my contributions to the defenses for the aforementioned attacks. Next, I will be presenting an introduction to IC reverse engineering (RE) attacks and the physical side-channel attacks (SCA).

\begin{figure}[!htb]
    \figSpace
    \centering
    \includegraphics[width=1\textwidth]{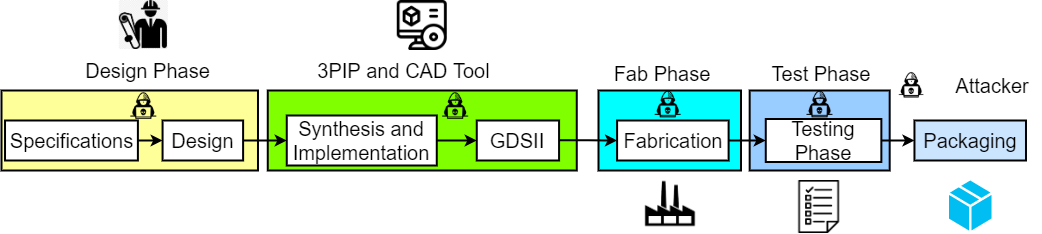}
    \caption{Integrated Chip (IC) supply chain process}
    \label{fig:supply_chain}
    \figSpace
\end{figure}

\subsubsection{IC Reverse Engineering Attacks}
IC manufacturing with millions and billions of transistors has become the need of the hour to support large consumer market that constantly demands increasing performance and speed. The semiconductor industry has progressed a lot in terms of speed and performance to satiate the needs but at the cost of manufacturing the IC offshore as a majority of companies have gone fabless due to various economic, cost and technology reasons thus exposing the IC supply chain to adversaries/attackers causing trust issues. The base reason for fabricating the design offshore is the smaller technology node, lower the number better the speed, it offers. The IC fabrication goes through a variety of design processes and each stage has become an access point where attackers can get their hands on to serve malicious purposes. 
The IC manufacturing chain is highly complex and is susceptible to various threats both from the outside and inside. Figure \ref{fig:supply_chain} presents the supply chain process, each process block is vulnerable to different type of attacks \cite{iscas_survey_dhavlle}. This has impacted the fabrication business economy due to the threats that are posed by untrusted fabs like reverse engineering, hardware trojans and IP piracy \cite{Rostami_IEEE'14, Torrance_dac'11, Tehranipoor_IEEE'17}. The IC design house imports the IP - intellectual property - from the IP vendors which are then integrated with the custom designs in the design house. The finalized design is thus sent to the Fabrication vendors for manufacturing. The hardware level attack could be executed at any of these stages. Logic locking \cite{Roy_date'08, Rajendran_ieee'15, Xie_ieee'19}, aka obfuscation, is a defense mechanism that is used to hide the true functionality of the circuit to prevent design leakage even if the attacker gets the design netlist. Additional gates are embedded in various parts of the design to hide the functionality and these gates only act as transparent via media when correct logic is supplied. The correct logic is known as keys which are supplied to the circuit during the activation phase and are stored in the on-chip tamper proof memory. The activated IC works as designed only after the correct combination of the keys. The keys is a chain of `1's and `0's which is only known to the designer. With the state-of-the-art attacks \cite{Rostami_IEEE'14, Torrance_dac'11, Tehranipoor_IEEE'17} prevalent, protection using only obfuscation technique does not suffice to thwart the attack. 
A recent state-of-the-art attack known as SAT attack is based on the assumption that the attacker has access to the functional IC (IC activated by the designer using the correct key combinations) and the locked/obfuscated netlist has surged. The SAT attack \cite{sat_attack_22} is based on the application of Distinguishing Input Patterns (DIPs) where the attack reduces key search space by iteratively applying DIPs to the obfuscated design and then applying the same DIPs to the functional IC thus eliminating incorrect key inputs. Along with logic locking type defense, split manufacturing \cite{Imeson_usenix'13} is a technique used to `split' and manufacture different blocks of the design at different fabrication units. As a consequence, an IC responsible for activation of secondary IC could be manufactured at a trusted foundry, while other IC, at smaller technology, can be manufactured offshore (potential untrusted foundry). After manufacturing the designer stacks the two components together so the trusted platform can activate and control its untrusted counterpart. To achieve this methodology, an application to facilitate and evaluate different logic locking techniques against SAT attack was indispensable. The developed application will be described in the next chapter. Also, apart from the application, I was entrusted to develop and test hardware communication modules responsible for secure data transfer between the trusted and untrusted platforms. The same is presented in the next chapter.

\subsubsection{Physical SCA}
Data integrity and security became an essential part in the era of digital 
systems where privacy and confidentiality needs to be ensured. There have been a plethora of works addressing the attacks on systems, like those posed by malware \cite{Sanket_ictai_2019, Sanket_icmla_2019, Sanket_dac'21, Dhavlle_date'21}, reverse engineering of hardware \cite{Gaurav_iccad'19, Gaurav_date'20}; attacks on machine-learning assisted hardware-based malware detectors (HMDs) \cite{SMPD_DAC'19, Sanket_esweek_2019, dhavlle_arxiv_21}, adversarial attacks on machine learning \cite{Sanket_glsvlsi'21}, machine learning based attacks on hardware \cite{Dhavlle_ahost'20, Dhavlle_Iscas'21}, cache based side-channel attacks \cite{Brasser_cases'18, Dhavlle_tcad'22, Abhijitt_isqed'20}, etc. Of these, side-channel attack and cryptosystem has been discussed in this work. 
Cyrptographic mechanisms are employed to offer security to the data by encrypting the data 
streams with a secret key and transform the data into a human non-readable 
format. The attempt to exercise a brute force to decrypt the information 
is exhaustive and can even be unfeasible. 
To efficiently decode the secret key and decrypt the information, 
adversaries target utilizing the information obtained through 
side-channels, termed as side-channel attacks. 
Side-channels are inherent in any given design and side-channel attacks exploit the information from these rather than exploiting vulnerabilities in the software. 
There exist both physical and microarchitectural side-channels that can leak secure critical information 
through acoustics, electromagnetic (EM) radiations, power trace, thermal maps and cache-access information. 
Power signature based side-channel threats are a pivotal threat as power consumption is an inherent and preliminary characteristic of any digital system. 

In this work I consider a power signature based side-channel attack on encryption algorithm 
executing on FPGAs as they
are proliferating into data centers for compute-intensive operations such as encryption. 
For the power analysis based SCA to be successful, the attacker measures the power traces from the system while triggering crypto operations on the system. 
This trace is then studied statistically to deduce the secret key. The fundamental principle underlying this attack is that different modules (operations) of AES consume different power, 
and thereby studying the power trace reveals the operation, 
based on which the secret key can be deduced.

Pengyuan Yu et al. in \cite{Yu_codes'07} propose an intelligent place-and-route technique to facilitate symmetrical routing as a defense against power analysis SCA on FPGA. Work in \cite{Yuval_crypto'03} describes how a circuit can be 
transformed to a larger circuit to defend against probe-based physical SCAs, but, the technique proposed is very complex. Work in \cite{Kocher_crypto'99} and \cite{Chari'99} describes algorithmic countermeasures to thwart SCAs which attempts to minimize the correlation between the intermediate values and the secret key ;and by algorithmically adding noise respectively. Also, circuit-level countermeasures are presented in papers \cite{Suzuki_IEICE'07,Trichina_IACR'03,Yang_DATE'05, Dubey_host'20, Matovu_dasp'20}. It is observed that the existing defenses require modifications in physical designs, leading to larger overheads and design complexity. 

To overcome these challenges and defend against power analysis SCA, I propose 
\shrt. More details about the proposed \shrt\ will be discussed in the coming chapters. 

%% file: chapterTwo.tex
\section[IC Reverse Engineering Defense]{IC Reverse Engineering Defense}
\subsection{Trusted-Untrusted Design Integration}
As discussed in the introduction section, a design could be subjected to IC reverse engineering (RE) attacks and design integrity attacks. Also, for security critical applications, the end design should be trustworthy. To address the IC manufacturing supply chain attacks \cite{iscas_survey_dhavlle, Rostami_IEEE'14, Torrance_dac'11, Tehranipoor_IEEE'17}, the idea of the project is to integrate two chips post manufacturing - one chip being a trusted chip and other could be manufactured at an offshore untrusted environment. The trusted environment could be a secure facility/foundry, and the untrusted could be a manufacturing unit outside the secure zone. To protect the untrusted design (UD) from attacks, the design would be obfuscated before the tapeout phase. The obfuscation \cite{Rajendran_ccs'13, Vijayakumar_tifs'17, Yasin_iccad'16,Roy_date'08, Rajendran_ieee'15, Xie_ieee'19} adds additional gates to the design that do not contribute to the end functionality, yet inserted to increase robustness against SAT-Solver or SAT attacks \cite{sat_attack_22}. The design is executed in different phases and evaluating robustness against attacks is one of the most important parts. Usually, obfuscation of the design is achieved by running various algorithms on a system, which is a laborious effort. Moreover, after the design framework is handed over to the end user, the user may not be familiar with the technicalities of the aforementioned obfuscation defenses and attacks. Automating the frontend tasks was proposed as a panacea to reduce the efforts in evaluating the attack and to render the system end-user friendly. Next, we will discuss more on the automation tool.

\begin{figure}[!htb]
   \figSpace
    \centering
    \includegraphics[width=1\textwidth]{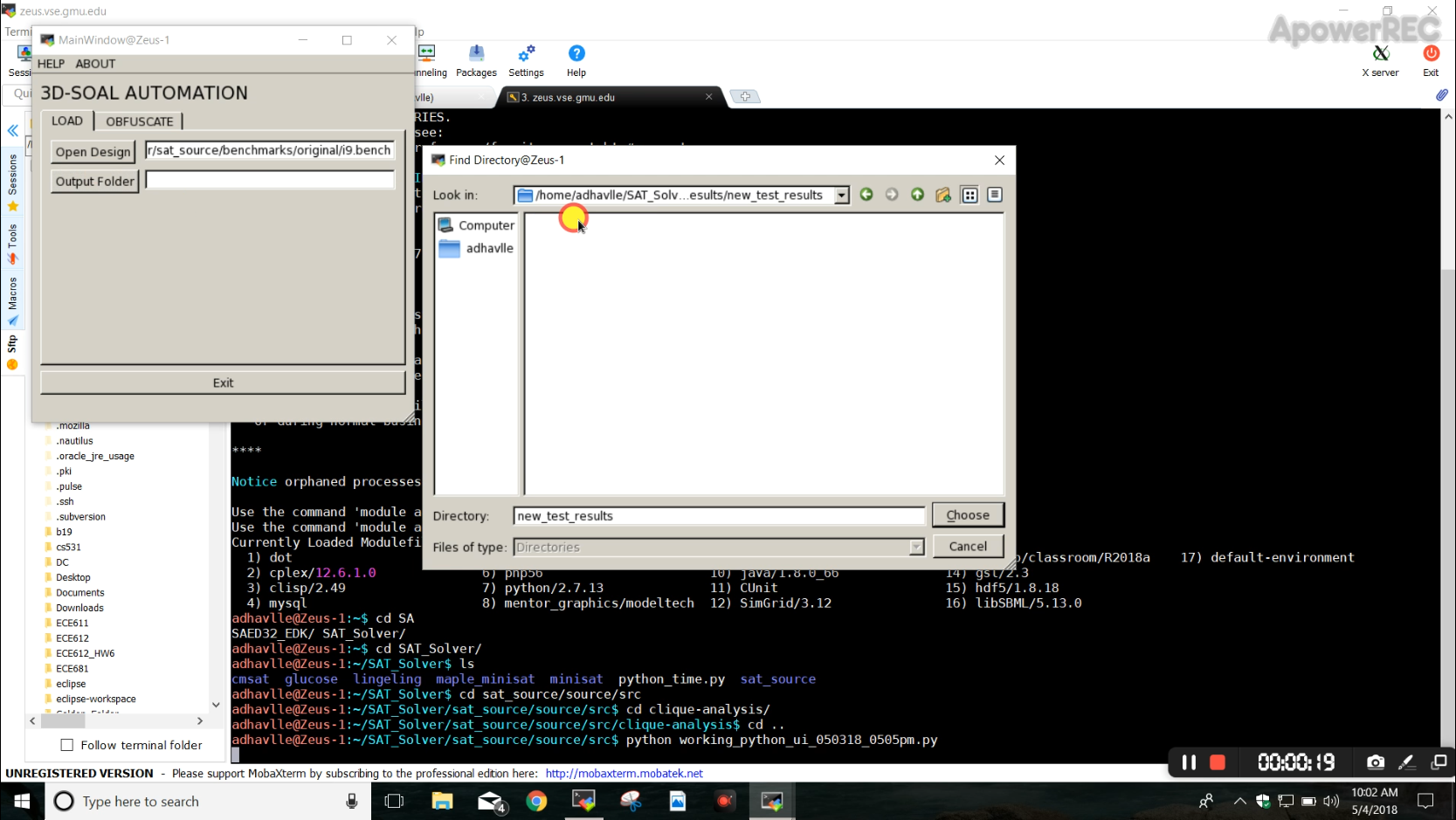}
    \caption{Choosing a design folder to select an input design}
    \label{fig:input}
    \figSpace
\end{figure}

\begin{figure}[!htb]
\figSpace
    \centering
    \includegraphics[width=1\textwidth]{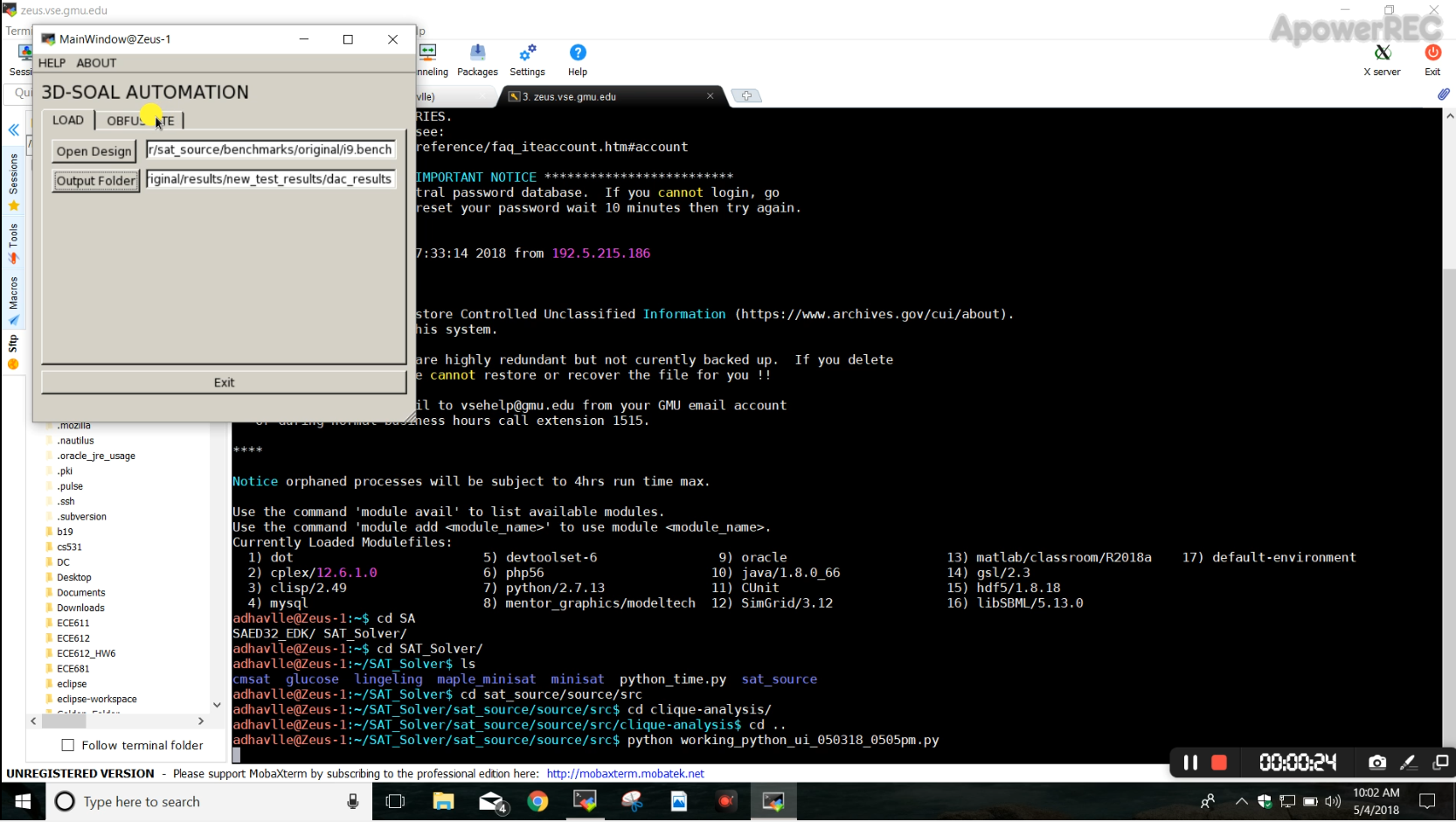}
    \caption{Choosing an output folder to save the resultant obfuscated design}
    \label{fig:output}
    \figSpace
\end{figure}

\subsection{Automation Tool for Design Security}
The automation application as shown in Figure \ref{fig:input} provides a user-friendly interface to obfuscate a design using various algorithms, and to de-obfuscate the same. Please note that the obfuscation algorithms are standard and I have no contributions towards them. My task was to build the application that could utilize the standard algorithms in the background. 

\begin{figure}[!htb]
\figSpace
    \centering
    \includegraphics[width=0.8\textwidth]{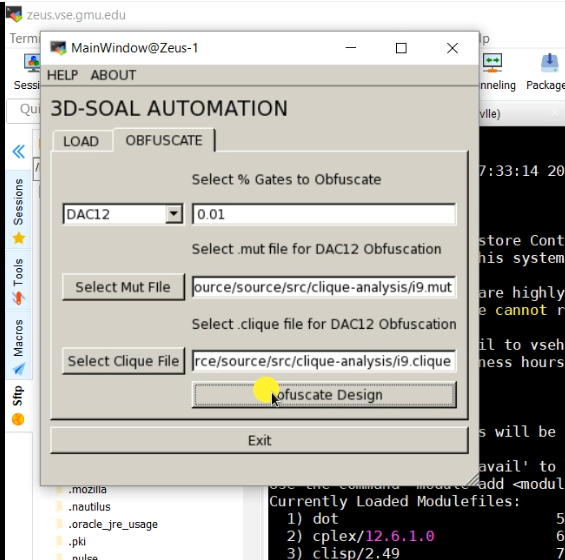}
    \caption{Selecting a obfuscation type, obfuscation percentage, and necessary files before obfuscating a design}
    \label{fig:select_type}
    \figSpace
\end{figure}

\begin{figure}[!htb]
\figSpace
    \centering
    \includegraphics[width=0.8\textwidth]{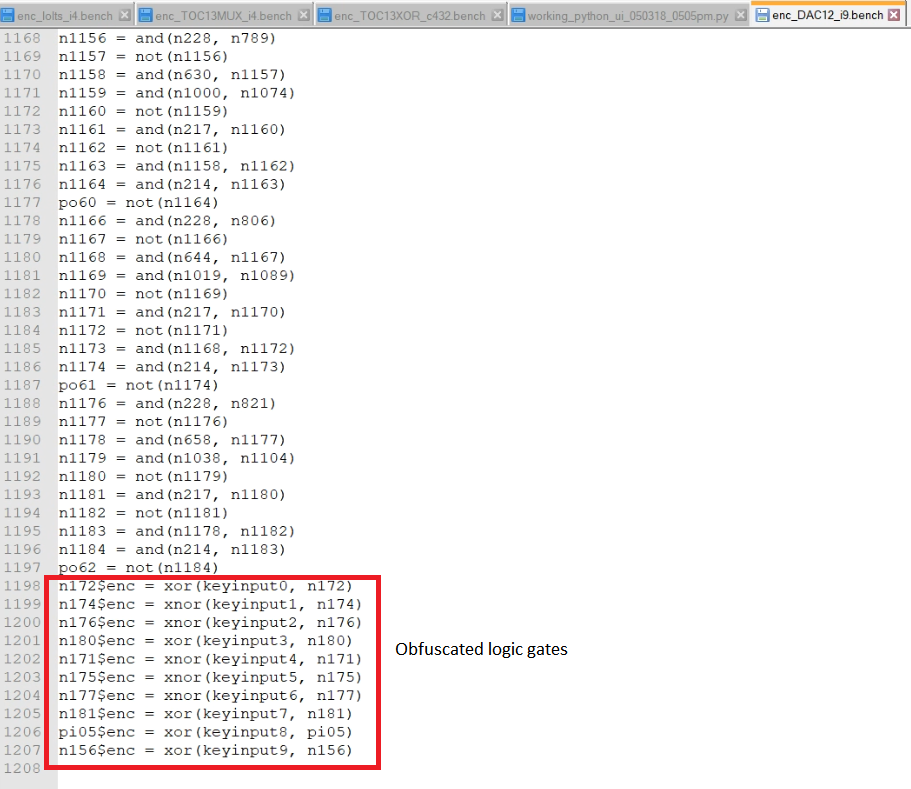}
    \caption{Output of an obfuscated design; the additional gates added at the end accommodate the secret key inputs} 
    \label{fig:obfuscated_output}
    \figSpace
\end{figure}

\begin{figure}[!htb]
\figSpace
    \centering
    \includegraphics[width=0.8\textwidth]{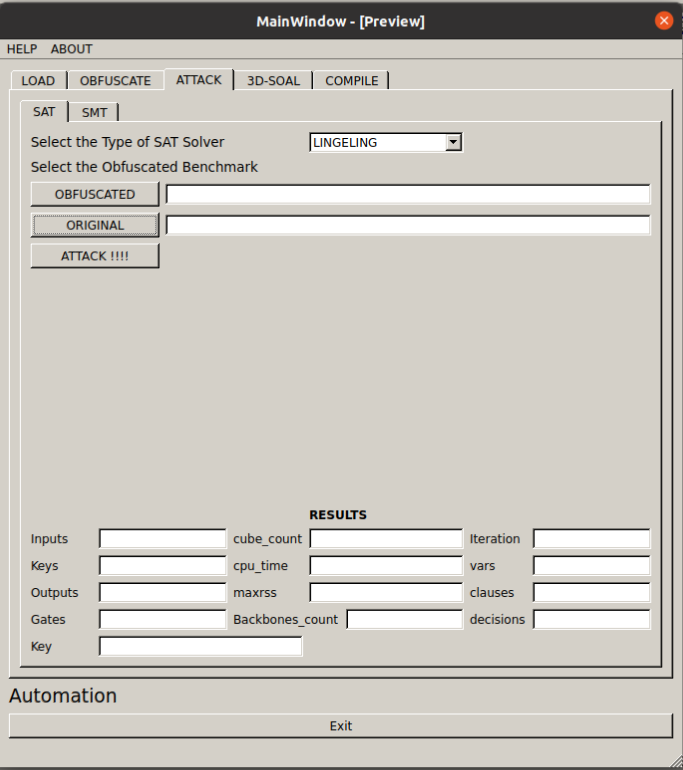}
    \caption{Automation for deploying different types of SAT attacks against obfuscated designs for robustness evaluation} 
    \label{fig:sat_attack}
    \figSpace
\end{figure}

The parameters the application is capable of selecting are listed below: 

\begin{enumerate}
    \item The obfuscation algorithms the application can use are:
    \begin{itemize}
    \item Iolts
    \item Random
    \item DAC12
    \item TOC13MUX
    \item TOC13XOR
\end{itemize}

\item The Attack tab has SAT and SMT type solvers. Within SAT, the application allows for five different solvers.
\begin{itemize}
    \item Lingeling
    \item CMSAT
    \item Glucose
    \item Maple\_minisat
    \item Minisat
\end{itemize}

\item   Number (in percent) of logic gates to obfuscate.

\item Selecting specific files required for DAC12 algorithm - .MUT and .CLIQUE files.

\item Select the input and output folders to choose a base design from and to store the resultant outputs.
\end{enumerate}

Prerequisites to executing the application are:
\begin{enumerate}
    \item Ensure SAT solvers are installed on the system, which includes the dependencies also.
    \item Ensure Python is installed.
    \item Install Qt designer - this is a GUI designing application that takes less efforts than directly writing the Python code for the GUI blocks.
    \item Install PyQt-4/5 - this is an API to support the GUI designing. 
    \item The code for recreating the application is provided in Listing \ref{lst:pyqt}; the user interface's XML code is provided in Listing \ref{lst:xml}.
\end{enumerate}

The steps to produce the obfuscated design is as follows:
\begin{enumerate}
    \item Use the Load tab in the application and select the base design to obfuscate. Refer to Figure \ref{fig:input}.
    \item Select the output folder to store the resultant obfuscated design. Refer to Figure \ref{fig:output}.
    \item Move to the Obfuscate tab of the application and choose the obfuscation algorithm, percent of gates to obfuscate, and selecting .MUT and .CLIQUE files, if DAC12 was chosen earlier. Refer to Figure \ref{fig:select_type}.
    \item Hit the `obfuscate design' button to start the task.
    \item Find the generated file in the output folder chosen earlier, and verify the change in the resultant design file. Refer to Figure \ref{fig:obfuscated_output}.
    \item For executing SAT-type attacks, click on the Attacks tab; select the type of solver, select an obfuscated design and original (unobfuscated) file; click the attack button to start the process. The results section shows different parameters of the outcome of the attack. These parameters are parsed by the script and shown in each of the boxes. Refer to Figure \ref{fig:sat_attack}.
    \end{enumerate}

\subsubsection{PyQt Code for Recreating a Basic Layout}
\begin{lstlisting}[language=Python, label={lst:pyqt}, caption= PyQt code for recreating a basic layout]
from PyQt4 import QtCore, QtGui
try:
    _fromUtf8 = QtCore.QString.fromUtf8
except AttributeError:
    def _fromUtf8(s):
        return s

try:
    _encoding = QtGui.QApplication.UnicodeUTF8
    def _translate(context, text, disambig):
        return QtGui.QApplication.translate(context, text, disambig, _encoding)
except AttributeError:
    def _translate(context, text, disambig):
        return QtGui.QApplication.translate(context, text, disambig)

class Ui_MainWindow(object):
    def setupUi(self, MainWindow):
        MainWindow.setObjectName(_fromUtf8("MainWindow"))
        MainWindow.resize(452, 724)
        self.centralwidget = QtGui.QWidget(MainWindow)
        self.centralwidget.setObjectName(_fromUtf8("centralwidget"))
        self.gridLayout = QtGui.QGridLayout(self.centralwidget)
        self.gridLayout.setObjectName(_fromUtf8("gridLayout"))
        self.label = QtGui.QLabel(self.centralwidget)
        font = QtGui.QFont()
        font.setPointSize(14)
        self.label.setFont(font)
        self.label.setObjectName(_fromUtf8("label"))
        self.gridLayout.addWidget(self.label, 0, 0, 1, 1)
        self.tabWidget = QtGui.QTabWidget(self.centralwidget)
        self.tabWidget.setObjectName(_fromUtf8("tabWidget"))
        self.tab = QtGui.QWidget()
        self.tab.setObjectName(_fromUtf8("tab"))
        self.formLayout = QtGui.QFormLayout(self.tab)
        self.formLayout.setObjectName(_fromUtf8("formLayout"))
        self.pushButton_3 = QtGui.QPushButton(self.tab)
        self.pushButton_3.setObjectName(_fromUtf8("pushButton_3"))
        self.pushButton_3.clicked.connect(self.fetch_file)
        self.formLayout.setWidget(0, QtGui.QFormLayout.LabelRole, self.pushButton_3)
        self.lineEdit = QtGui.QLineEdit(self.tab)
        self.lineEdit.setObjectName(_fromUtf8("lineEdit"))
        self.formLayout.setWidget(0, QtGui.QFormLayout.FieldRole, self.lineEdit)
        self.pushButton_4 = QtGui.QPushButton(self.tab)
        self.pushButton_4.setObjectName(_fromUtf8("pushButton_4"))
        self.pushButton_3.clicked.connect(self.fetch_output_folder)
        self.formLayout.setWidget(1, QtGui.QFormLayout.LabelRole, self.pushButton_4)
        self.lineEdit_3 = QtGui.QLineEdit(self.tab)
        self.lineEdit_3.setObjectName(_fromUtf8("lineEdit_3"))
        self.formLayout.setWidget(1, QtGui.QFormLayout.FieldRole, self.lineEdit_3)
        self.tabWidget.addTab(self.tab, _fromUtf8(""))
        self.tab_2 = QtGui.QWidget()
        self.tab_2.setObjectName(_fromUtf8("tab_2"))
        self.formLayout_2 = QtGui.QFormLayout(self.tab_2)
        self.formLayout_2.setFieldGrowthPolicy(QtGui.QFormLayout.AllNonFixedFieldsGrow)
        self.formLayout_2.setObjectName(_fromUtf8("formLayout_2"))
        self.label_4 = QtGui.QLabel(self.tab_2)
        self.label_4.setObjectName(_fromUtf8("label_4"))
        self.formLayout_2.setWidget(0, QtGui.QFormLayout.FieldRole, self.label_4)
        self.comboBox = QtGui.QComboBox(self.tab_2)
        self.comboBox.setObjectName(_fromUtf8("comboBox"))
        self.comboBox.addItem(_fromUtf8(""))
        self.comboBox.addItem(_fromUtf8(""))
        self.comboBox.addItem(_fromUtf8(""))
        self.comboBox.addItem(_fromUtf8(""))
        self.comboBox.addItem(_fromUtf8(""))
        self.formLayout_2.setWidget(1, QtGui.QFormLayout.LabelRole, self.comboBox)
        self.lineEdit_2 = QtGui.QLineEdit(self.tab_2)
        self.lineEdit_2.setObjectName(_fromUtf8("lineEdit_2"))
        self.formLayout_2.setWidget(1, QtGui.QFormLayout.FieldRole, self.lineEdit_2)
        self.pushButton_2 = QtGui.QPushButton(self.tab_2)
        self.pushButton_2.setObjectName(_fromUtf8("pushButton_2"))
        self.formLayout_2.setWidget(2, QtGui.QFormLayout.FieldRole, self.pushButton_2)
        self.tabWidget.addTab(self.tab_2, _fromUtf8(""))
        self.gridLayout.addWidget(self.tabWidget, 1, 0, 1, 1)
        self.pushButton = QtGui.QPushButton(self.centralwidget)
        self.pushButton.setObjectName(_fromUtf8("pushButton"))
        self.gridLayout.addWidget(self.pushButton, 2, 0, 1, 1)
        self.label_2 = QtGui.QLabel(self.centralwidget)
        self.label_2.setObjectName(_fromUtf8("label_2"))
        self.gridLayout.addWidget(self.label_2, 3, 1, 1, 1)
        MainWindow.setCentralWidget(self.centralwidget)
        self.statusbar = QtGui.QStatusBar(MainWindow)
        self.statusbar.setObjectName(_fromUtf8("statusbar"))
        MainWindow.setStatusBar(self.statusbar)
        self.menubar = QtGui.QMenuBar(MainWindow)
        self.menubar.setGeometry(QtCore.QRect(0, 0, 452, 26))
        self.menubar.setObjectName(_fromUtf8("menubar"))
        self.menuHELP = QtGui.QMenu(self.menubar)
        self.menuHELP.setObjectName(_fromUtf8("menuHELP"))
        self.menuABOUT = QtGui.QMenu(self.menubar)
        self.menuABOUT.setObjectName(_fromUtf8("menuABOUT"))
        MainWindow.setMenuBar(self.menubar)
        self.actionDont_ask_for_any_help = QtGui.QAction(MainWindow)
        self.actionDont_ask_for_any_help.setObjectName(_fromUtf8("actionDont_ask_for_any_help"))
        self.actionHkjhk = QtGui.QAction(MainWindow)
        self.actionHkjhk.setObjectName(_fromUtf8("actionHkjhk"))
        self.actionAbout_3D_SOAL_Automator = QtGui.QAction(MainWindow)
        self.actionAbout_3D_SOAL_Automator.setObjectName(_fromUtf8("actionAbout_3D_SOAL_Automator"))
        self.menuHELP.addSeparator()
        self.menuHELP.addAction(self.actionDont_ask_for_any_help)
        self.menuABOUT.addAction(self.actionAbout_3D_SOAL_Automator)
        self.menubar.addAction(self.menuHELP.menuAction())
        self.menubar.addAction(self.menuABOUT.menuAction())

        self.retranslateUi(MainWindow)
        self.tabWidget.setCurrentIndex(0)
        QtCore.QMetaObject.connectSlotsByName(MainWindow)

    def retranslateUi(self, MainWindow):
        MainWindow.setWindowTitle(_translate("MainWindow", "3D-SOAL Automator", None))
        self.label.setText(_translate("MainWindow", "3D-SOAL AUTOMATION", None))
        self.pushButton_3.setText(_translate("MainWindow", "Open Design", None))
        self.pushButton_4.setText(_translate("MainWindow", "Output Folder", None))
        self.tabWidget.setTabText(self.tabWidget.indexOf(self.tab), _translate("MainWindow", "LOAD", None))
        self.label_4.setText(_translate("MainWindow", "Select % Gates to Obfuscate", None))
        self.comboBox.setItemText(0, _translate("MainWindow", "Iolts", None))
        self.comboBox.setItemText(1, _translate("MainWindow", "Random", None))
        self.comboBox.setItemText(2, _translate("MainWindow", "DAC", None))
        self.comboBox.setItemText(3, _translate("MainWindow", "MUX", None))
        self.comboBox.setItemText(4, _translate("MainWindow", "TOC13", None))
        self.pushButton_2.setText(_translate("MainWindow", "Obfuscate Design", None))
        self.tabWidget.setTabText(self.tabWidget.indexOf(self.tab_2), _translate("MainWindow", "OBFUSCATE", None))
        self.pushButton.setText(_translate("MainWindow", "Exit", None))
        self.label_2.setText(_translate("MainWindow", "Created By :SSS, AAD", None))
        self.menuHELP.setTitle(_translate("MainWindow", "HELP", None))
        self.menuABOUT.setTitle(_translate("MainWindow", "ABOUT", None))
        self.actionDont_ask_for_any_help.setText(_translate("MainWindow", "Dont ask for any help!!", None))
        self.actionHkjhk.setText(_translate("MainWindow", "hkjhk", None))
        self.actionAbout_3D_SOAL_Automator.setText(_translate("MainWindow", "About 3D-SOAL Automator", None))

    def fetch_file(self):
        dlg = QFileDialog()
        dlg.setFileMode(QFileDialog.AnyFile)
        dlg.setFilter("Benchmark File (*.bench)")
        filenames = QStringList()
        self.lineEdit.setText(filenames)

    def fetch_output_folder(self):
        dlg1 = QFileDialog()
        dlg1.setFileMode(QFileDialog.AnyFile)
        #dlg1.setFilter("Benchmark File (*.bench)")
        folder_names = QStringList()
        self.lineEdit3.setText(folder_names)

if __name__ == "__main__":
    import sys
    app = QtGui.QApplication(sys.argv)
    MainWindow = QtGui.QMainWindow()
    ui = Ui_MainWindow()
    ui.setupUi(MainWindow)
    MainWindow.show()
    sys.exit(app.exec_())
\end{lstlisting}
\vspace{5em}
\subsubsection{User Interface (UI) designed in QT Designer}
\begin{lstlisting}[language=XML, label={lst:xml},caption= UI code designed in QT designer for the application]
<?xml version="1.0" encoding="UTF-8"?>
<ui version="4.0">
 <class>MainWindow</class>
 <widget class="QMainWindow" name="MainWindow">
  <property name="geometry">
   <rect>
    <x>0</x>
    <y>0</y>
    <width>452</width>
    <height>724</height>
   </rect>
  </property>
  <property name="windowTitle">
   <string>MainWindow</string>
  </property>
  <widget class="QWidget" name="centralwidget">
   <layout class="QGridLayout" name="gridLayout">
    <item row="0" column="0">
     <widget class="QLabel" name="label">
      <property name="font">
       <font>
        <pointsize>14</pointsize>
       </font>
      </property>
      <property name="text">
       <string>3D-SOAL AUTOMATION</string>
      </property>
     </widget>
    </item>
    <item row="1" column="0">
     <widget class="QTabWidget" name="tabWidget">
      <property name="currentIndex">
       <number>1</number>
      </property>
      <widget class="QWidget" name="tab">
       <attribute name="title">
        <string>LOAD</string>
       </attribute>
       <layout class="QFormLayout" name="formLayout">
        <item row="0" column="0">
         <widget class="QPushButton" name="pushButton_3">
          <property name="text">
           <string>Open Design</string>
          </property>
         </widget>
        </item>
        <item row="0" column="1">
         <widget class="QLineEdit" name="lineEdit"/>
        </item>
        <item row="1" column="0">
         <widget class="QPushButton" name="pushButton_4">
          <property name="text">
           <string>Output Folder</string>
          </property>
         </widget>
        </item>
        <item row="1" column="1">
         <widget class="QLineEdit" name="lineEdit_3"/>
        </item>
       </layout>
      </widget>
      <widget class="QWidget" name="tab_2">
       <attribute name="title">
        <string>OBFUSCATE</string>
       </attribute>
       <layout class="QFormLayout" name="formLayout_2">
        <property name="fieldGrowthPolicy">
         <enum>QFormLayout::AllNonFixedFieldsGrow</enum>
        </property>
        <item row="0" column="1">
         <widget class="QLabel" name="label_4">
          <property name="text">
           <string>Select % Gates to Obfuscate</string>
          </property>
         </widget>
        </item>
        <item row="1" column="0">
         <widget class="QComboBox" name="comboBox">
          <item>
           <property name="text">
            <string>Iolts</string>
           </property>
          </item>
          <item>
           <property name="text">
            <string>Random</string>
           </property>
          </item>
          <item>
           <property name="text">
            <string>DAC</string>
           </property>
          </item>
          <item>
           <property name="text">
            <string>MUX</string>
           </property>
          </item>
          <item>
           <property name="text">
            <string>TOC13</string>
           </property>
          </item>
         </widget>
        </item>
        <item row="1" column="1">
         <widget class="QLineEdit" name="lineEdit_2"/>
        </item>
        <item row="2" column="1">
         <widget class="QPushButton" name="pushButton_2">
          <property name="text">
           <string>Obfuscate Design</string>
          </property>
         </widget>
        </item>
       </layout>
      </widget>
     </widget>
    </item>
    <item row="2" column="0">
     <widget class="QPushButton" name="pushButton">
      <property name="text">
       <string>Exit</string>
      </property>
     </widget>
    </item>
    <item row="3" column="1">
     <widget class="QLabel" name="label_2">
      <property name="text">
       <string>Created By :SSS, AAD</string>
      </property>
     </widget>
    </item>
   </layout>
  </widget>
  <widget class="QStatusBar" name="statusbar"/>
  <widget class="QMenuBar" name="menubar">
   <property name="geometry">
    <rect>
     <x>0</x>
     <y>0</y>
     <width>452</width>
     <height>26</height>
    </rect>
   </property>
   <widget class="QMenu" name="menuHELP">
    <property name="title">
     <string>HELP</string>
    </property>
    <addaction name="separator"/>
    <addaction name="actionDont_ask_for_any_help"/>
   </widget>
   <widget class="QMenu" name="menuABOUT">
    <property name="title">
     <string>ABOUT</string>
    </property>
    <addaction name="actionAbout_3D_SOAL_Automator"/>
   </widget>
   <addaction name="menuHELP"/>
   <addaction name="menuABOUT"/>
  </widget>
  <action name="actionDont_ask_for_any_help">
   <property name="text">
    <string>Dont ask for any help!!</string>
   </property>
  </action>
  <action name="actionHkjhk">
   <property name="text">
    <string>hkjhk</string>
   </property>
  </action>
  <action name="actionAbout_3D_SOAL_Automator">
   <property name="text">
    <string>About 3D-SOAL Automator</string>
   </property>
  </action>
 </widget>
 <resources/>
 <connections/>
</ui>
\end{lstlisting}

\subsection{Reconfigurable Modules for Trusted-Untrusted Platform Communication}
\begin{figure}[!htb]
\figSpace
    \centering
    \includegraphics[width=1\textwidth]{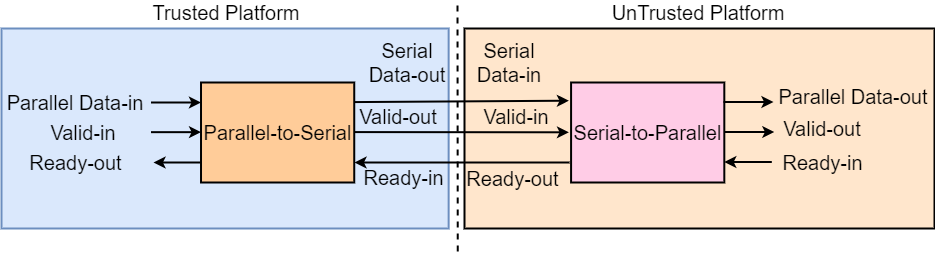}
    \caption{Hardware communication modules that establish communication between the trusted and untrusted platforms}
    \label{fig:comm_modules}
    \figSpace
\end{figure}

\begin{figure}
\figSpace
    \centering
    \includegraphics[width=0.8\textwidth]{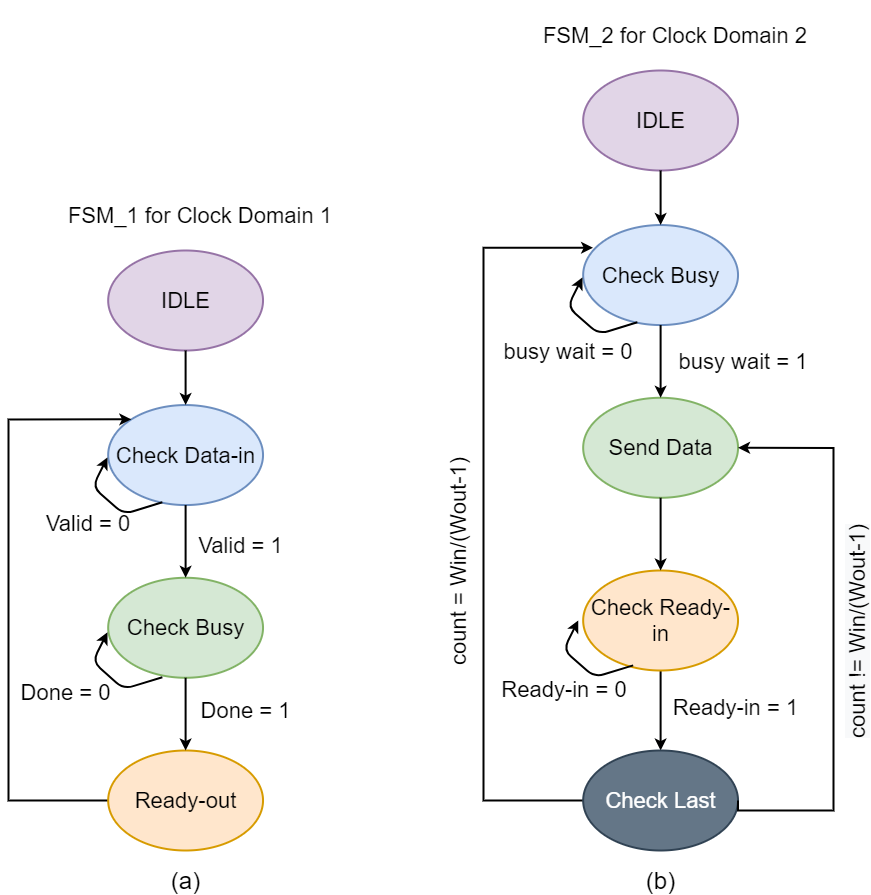}
    \caption{Finite State Machine (FSM) diagrams for input and output side, which utilize two different clocks}
    \label{fig:fsm}
    \figSpace
\end{figure}

The trusted and untrusted platforms are stacked/connected together so the master platform can control the other untrusted platform. Depending on the scenario and other design constraints there may arise a need to either establish a serial (less wires and slower data) communication versus a parallel (more wires and faster data) communication, or vice versa. To address this, I was asked to design and test a parallel-to-parallel hardware communication modules. The interface of both the platforms is a serial connection, the final ends being a parallel data-in and data-out approach. The two platforms may have different operating clock speeds. This could also be referred as a clock domain crossing scenario. Hence, it is necessary to realize the design using finite state machine (FSM) approach. The block diagram of the communication blocks is shown in Figure \ref{fig:comm_modules}; the block diagrams of the state machine is shown in Figure \ref{fig:fsm}. Referring to Figure \ref{fig:comm_modules}, the modules use a AXI interface to communicate with each other and other blocks of the higher level design. The AXI interface has standardized signal names to denote their functionality. The data bus could be an input or a output that takes data into the block or out from the block. The valid signal denotes that the content on the data bus is `valid' and ready to be consumed. The valid could be an input or an output signal depending on whether it is placed on the consumer side or the generator side. A consumer block accepts the data from the block previous to it; a generator block generates some data and feeds it to the consumer block after it. A ready signal indicates the block is ready to process next piece of data.

The two state machines with different clocks is shown in Figure \ref{fig:fsm}. The state machine begins in Idle state; part (a) waits for the valid signal and content on the data bus. The Check Busy state checks if the done signal goes high, after which the ready-out signal is set to `1'. The done signal is active when the data is processed by the block and ready to be transferred to the next block. In part (b), the Check Busy state expects a busy wait signal based on which it moves to the next state, Send Data, which converts the parallel data into serial data. While this conversion occurs, the state machine checks for the last piece of data to be processed in the Check Last state. After the conversion of data, the Send Data block will transfer the data out on the bus if the Ready-in signal is active.

\begin{figure}
\figSpace
    \centering
    \includegraphics[width=1\textwidth]{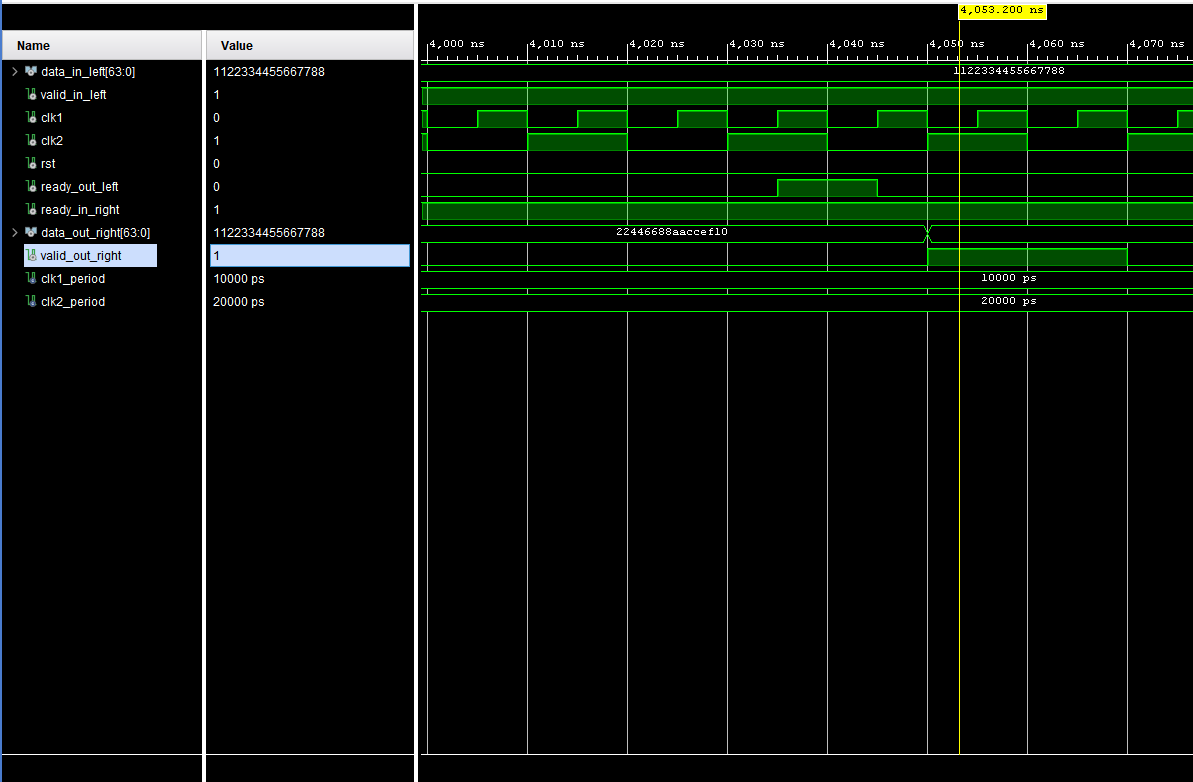}
    \caption{Simulation waveform for 64-bit parallel input}
    \label{fig:input_64}
    \figSpace
\end{figure}

\begin{figure}
\figSpace
    \centering
    \includegraphics[width=1\textwidth]{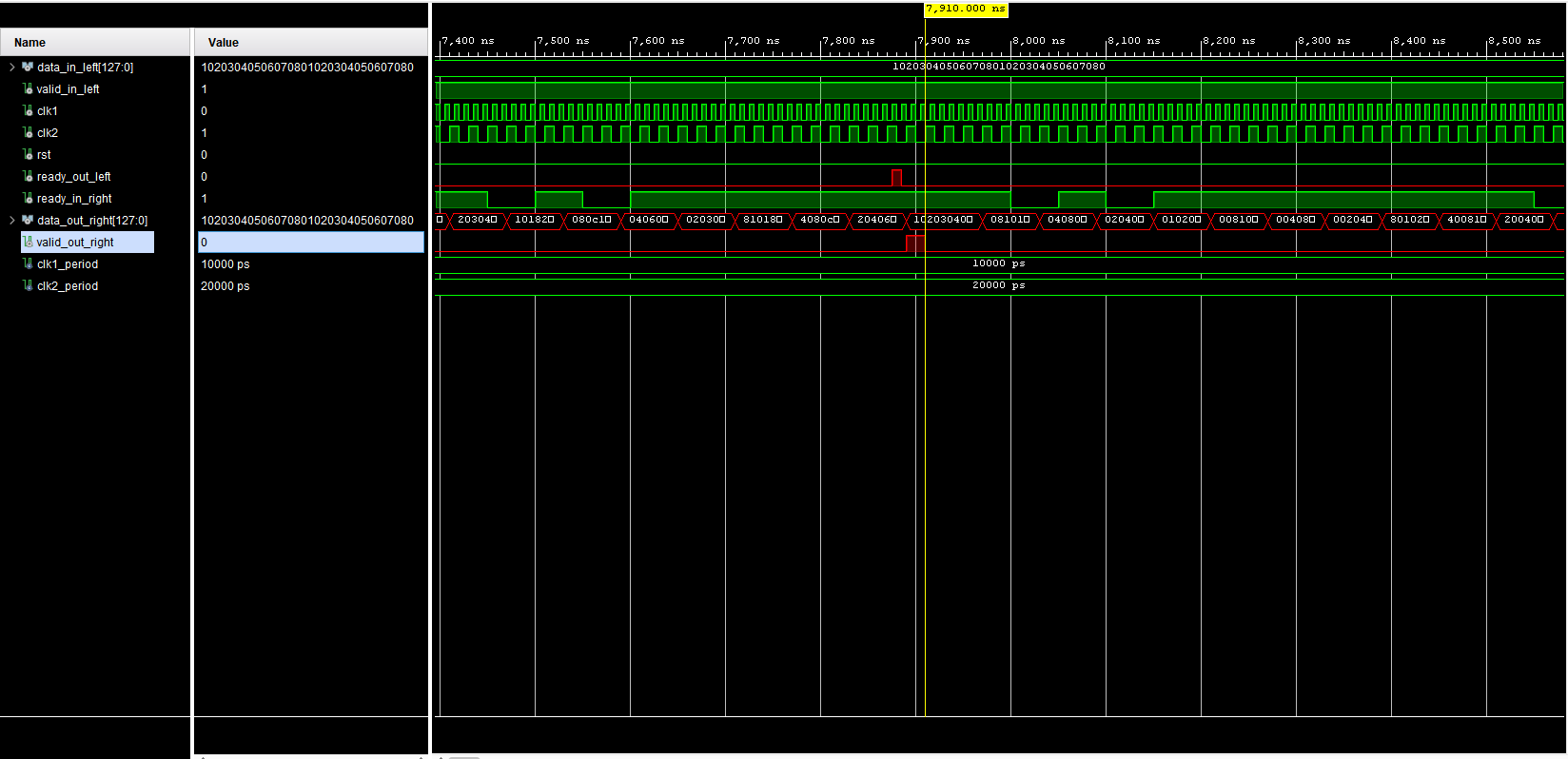}
    \caption{Simulation waveform for 128-bit parallel input}
    \label{fig:input_128}
    \figSpace
\end{figure}

\begin{figure}
\figSpace
    \centering
    \includegraphics[width=1\textwidth]{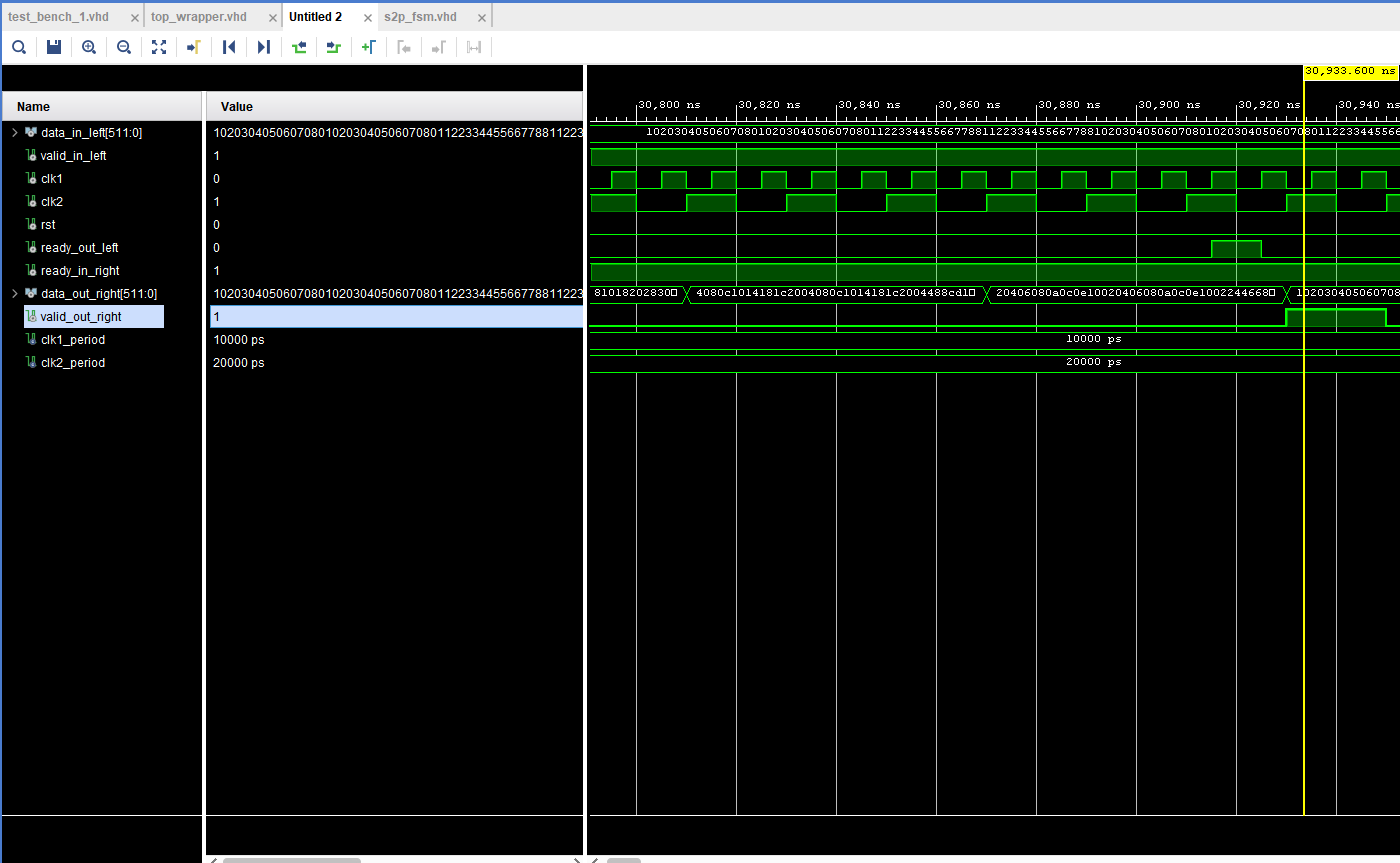}
    \caption{Simulation waveform for 512-bit parallel input}
    \label{fig:input_512}
    \figSpace
\end{figure}
The modules were realized in VHDL language and Xilinx Vivado was used to simulate, synthesize and generate waveforms. The simulation waveforms for 64, 128, and 512 bit data input are shown in Figure \ref{fig:input_64}, \ref{fig:input_128}, and \ref{fig:input_512}. It can be seen in the diagrams that the ready-out signal is set to `1' when the data processing is complete. The code for the parallel to serial hardware module is presented in Listing \ref{lst:p2s}. The resource utilization for 64, 128 and 512-bits configuration implemented on a Zedboard FPGA is shown in Figure \ref{tbl:overhead_64_bits}, \ref{tbl:overhead_128_bits}, and \ref{tbl:overhead_512_bits}.

\subsubsection{VHDL code for Parallel to Serial Module}
\begin{lstlisting}[language=VHDL, label={lst:p2s}, caption= VHDL code for realizing the parallel to serial converter hardware module]

library IEEE;
use IEEE.STD_LOGIC_1164.ALL;
use IEEE.STD_LOGIC_SIGNED.ALL;
use IEEE.numeric_std.all;
use work.my_package.all;

entity P2S is
    generic(
        w_out : integer := 1;
        w_in : integer := 512
    );
    Port ( data_in : in  STD_LOGIC_VECTOR (w_in-1 downto 0);
           valid_in : in  STD_LOGIC;
           ready_out : out  STD_LOGIC;
           clk1 : in  STD_LOGIC;
           clk2 : in  STD_LOGIC;
           rst  : in  STD_LOGIC;
           data_out : out  STD_LOGIC_VECTOR (w_out-1 downto 0);
           valid_out : out  STD_LOGIC;
           ready_in : in  STD_LOGIC);
end P2S;

architecture Behavioral of P2S is

type state_type1 is (FSM1_IDLE, FSM1_CHKDIN, FSM1_CHKBUSY, FSM1_READYOUT);
type state_type2 is (FSM2_IDLE, FSM2_CHKBUSY, FSM2_SEND_DATA, FSM2_CHK_READYIN, FSM2_CHKLAST, FSM2_DONE); 

type T_SLVV is array (NATURAL range <>)of std_logic_vector(w_out-1 downto 0);
signal state1, next_state1 : state_type1; 
signal state2, next_state2 : state_type2; 

--Declare internal signals for all outputs of the state-machine

signal busy_wait	: std_logic;  -- example output signal
signal done			: std_logic;  -- example output signal
signal count: std_logic_vector(log2(w_in/w_out)-1 downto 0);
signal mux_in : T_SLVV (w_in/w_out-1 downto 0);

signal ready_out_buff	: std_logic;  -- example output signal
signal valid_out_buff	: std_logic;  -- example output signal

--other outputs
begin
valid_out <= valid_out_buff;
ready_out <= ready_out_buff;

--FSM1 defined 
FSM1_state_update: process (clk1)
   begin
      if (clk1'event and clk1 = '1') then
         if (rst = '1') then
            state1 <= FSM1_IDLE;
         else
            state1 <= next_state1;
         end if;        
      end if;
   end process;

FSM1_ready_out: process (state1, rst, clk1, done, valid_in, count)
   begin
      if (state1 = FSM1_IDLE) then
         ready_out_buff <= '1';
      elsif(state1 = FSM1_READYOUT) then
        ready_out_buff <= '1';
		
		elsif(state1 = FSM1_CHKDIN and valid_in = '1') then
			ready_out_buff <= '0';
      else
			ready_out_buff <= ready_out_buff;
		end if;

   end process;
	
FSM1_busy_wait: process (state1, rst, clk1, done, valid_in)
   begin
      if (state1 = FSM1_CHKDIN and valid_in = '1') then
         busy_wait <= '1';
      elsif(state1 = FSM1_READYOUT) then
         busy_wait <= '0';
      elsif(state1 = FSM1_IDLE) then
			busy_wait <= '0';
		else
			busy_wait <= busy_wait;
		end if;
   end process;
 
FSM1_digram: process (state1, rst, clk1, done, valid_in)
   begin
      next_state1 <= state1;
      case (state1) is
         when FSM1_IDLE =>
				next_state1 <= FSM1_CHKDIN;
         when FSM1_CHKDIN =>
            if valid_in = '1' then
               next_state1 <= FSM1_CHKBUSY;
				else
					next_state1 <= FSM1_CHKDIN;
            end if;
         when FSM1_CHKBUSY =>
				if done = '1' then
					next_state1 <= FSM1_READYOUT;
				else
					next_state1 <= FSM1_CHKBUSY;
				end if;
         when FSM1_READYOUT =>
				next_state1 <= FSM1_CHKDIN;
         when others =>
            next_state1 <= FSM1_IDLE;
      end case;      
   end process;

--FSM2 defined
FSM2_state_update: process (clk2)
   begin
      if (clk2'event and clk2 = '1') then
         if (rst = '1') then
            state2 <= FSM2_IDLE;
         else
            state2 <= next_state2;
         end if;        
      end if;
   end process;

FSM2_done_proc: process (state2, rst, clk2, count)
   begin
      if (state2 = FSM2_CHKLAST and count = w_in/w_out -1) then
         done <= '1';
      elsif (state2 = FSM2_CHKBUSY or state2 = FSM2_IDLE) then
			done <= '0';
		else
			done <= done;
		end if;
   end process;

FSM2_valid_out: process (state2, rst, clk2, count, busy_wait)
   begin
      if (state2 = FSM2_SEND_DATA and busy_wait = '1') then
         valid_out_buff <= '1';
		elsif (state2 = FSM2_CHK_READYIN) then
			valid_out_buff <= '0';
		elsif (state2 = FSM2_CHKLAST and count /= w_in/w_out -1) then 
			valid_out_buff <= '0';
      elsif (state2 = FSM2_CHKBUSY or state2 = FSM2_IDLE) then
			valid_out_buff <= '0';
		else
			valid_out_buff <= valid_out_buff;
		end if;
   end process;

FSM2_digram: process (state2, rst, clk2, busy_wait, ready_in, count)
   begin
      next_state2 <= state2;
      case (state2) is
         when FSM2_IDLE =>
				next_state2 <= FSM2_CHKBUSY;
         when FSM2_CHKBUSY =>
            if busy_wait = '0' then
               next_state2 <= FSM2_CHKBUSY;
				else
					next_state2 <= FSM2_SEND_DATA;
            end if;
         when FSM2_SEND_DATA =>
					next_state2 <= FSM2_CHK_READYIN;
         when FSM2_CHK_READYIN =>
				if ready_in = '0' then
					next_state2 <= FSM2_CHK_READYIN;
				else
					next_state2 <= FSM2_CHKLAST;
				end if;
         when FSM2_CHKLAST =>
				if count = w_in/w_out-1 then 
					next_state2 <= FSM2_CHKBUSY;
				else
					next_state2 <= FSM2_SEND_DATA;
				end if;
			when others =>
            next_state2 <= FSM2_IDLE;
      end case;      
   end process;

--process for counter
Counter: process (clk2) 
	begin
		if clk2='1' and clk2'event then
			if (rst = '1') then
				count <= (others => '0');
			elsif state2 = FSM2_CHKLAST then
				count <= count + 1;
			elsif(state2 = FSM2_CHKBUSY)then
				count <= (others => '0');
			elsif (state2 = FSM2_IDLE) then
				count <= (others => '0');
			else
				count <= count;
			end if;
		end if;
end process;
 
 --original one commented below 
	generate_data_out : for i in 0 to (w_in/w_out)-1 generate
        mux_in(i) <= data_in((I*w_out)+w_out-1 downto I*w_out);
   end generate; 
   data_out <= mux_in(to_integer(unsigned(count)));   
	
end Behavioral;

\end{lstlisting}

\begin{table}
\centering
\tableSpace
\caption{Resource utilization for P2S and S2P module for 64-bit parallel input data }
\label{tbl:overhead_64_bits}
\scalebox{0.7}{
\begin{tabular}{|l|l|l|l|l|l|l|}
\hline
\multirow{2}{7em}{Total Resource} & Slice LUTs & Slice Registers & F7 Muxes & F8 Muxes & Bonded IOB & BUFCTRL \\
\cline{2-7}
 & 53200 & 106400 & 26600 & 13300 & 200 & 32 \\
\hline
\multirow{2}{10em}{Top Wrapper Utilization (\%)} & 0.10 & 0.09 & 0.03 & 0.03 & 67.50 & 6.25 \\
 &&&&&& \\
\hline
\multirow{2}{10em}{P2S Utilization (\%)} & 0.07 & 0.02 & 0.03 & 0.03 & 0.0 & 0.0 \\
 &&&&&& \\
\hline

\multirow{2}{10em}{S2P Utilization (\%)} & 0.03 & 0.08 & 0.0 & 0.0 & 0.0 & 0.0 \\
 &&&&&& \\
\hline
\end{tabular}}
\tableSpace
\end{table}

\begin{table}[]
\centering
\tableSpace
\caption{Resource utilization for P2S and S2P module for 128-bit parallel input data }
\label{tbl:overhead_128_bits}
\scalebox{0.8}{
\begin{tabular}{|l|l|l|l|l|}
\hline
\multirow{2}{7em}{Total Resource} & Slice LUTs & Slice Registers & F7 Muxes & F8 Muxes \\
\cline{2-5}
 & 53200 & 106400 & 26600 & 13300\\
\hline
\multirow{2}{10em}{Top Wrapper Utilization (\%)} & 0.15 & 0.16 & 0.06 & 0.06 \\
 &&&& \\
\hline
\multirow{2}{10em}{P2S Utilization (\%)} & 0.11 & 0.02 & 0.06 & 0.06\\
 &&&& \\
\hline

\multirow{2}{10em}{S2P Utilization (\%)} & 0.04 & 0.14 & 0.0 & 0.0\\
 &&&& \\
\hline
\end{tabular}}
\tableSpace
\end{table}

\begin{table}[]
\centering
\tableSpace
\caption{Resource utilization for P2S and S2P module for 512-bit parallel input data }
\label{tbl:overhead_512_bits}
\scalebox{0.7}{
\begin{tabular}{|l|l|l|l|l|l|l|}
\hline
\multirow{2}{7em}{Total Resource} & Slice LUTs & Slice Registers & F7 Muxes & F8 Muxes & Bonded IOB & BUFCTRL \\
\cline{2-7}
 & 53200 & 106400 & 26600 & 13300 & 200 & 32 \\
\hline
\multirow{2}{10em}{Top Wrapper Utilization (\%)} & 0.35 & 0.52 & 0.26 & 0.24 & 515.50 & 6.25 \\
 &&&&&& \\
\hline
\multirow{2}{10em}{P2S Utilization (\%)} & 0.30 & 0.02 & 0.26 & 0.24 & 0.0 & 0.0 \\
 &&&&&& \\
\hline

\multirow{2}{10em}{S2P Utilization (\%)} & 0.04 & 0.50 & 0.0 & 0.0 & 0.0 & 0.0 \\
 &&&&&& \\
\hline
\end{tabular}}
\tableSpace
\end{table}

%% file: chapterThree.tex
\section[Defense Against CPA-based Physical Side-Channel Attack]{Defense Against CPA-based Physical Side-Channel Attack}
\subsection{Physical Side-Channel Attack}
In this section, an introduction to physical side-channel attacks is provided, followed by the setup used for power measurement and a background of correlation power analysis  (CPA). 
Data integrity and security became an essential part in the era of digital 
systems where privacy and confidentiality needs to be ensured. There have been a plethora of works addressing the attacks on systems, like those posed by malware \cite{Sanket_ictai_2019, Sanket_icmla_2019, Sanket_dac'21, Dhavlle_date'21}, reverse engineering of hardware \cite{Gaurav_iccad'19, Gaurav_date'20}, attacks on machine-learning assisted hardware-based malware detectors (HMDs) \cite{SMPD_DAC'19, Sanket_esweek_2019}, adversarial attacks on machine learning \cite{Sanket_glsvlsi'21}, cache based side-channel attacks \cite{Brasser_cases'18, Abhijitt_cases'19, Abhijitt_isqed'20}, etc. Of these, side-channel attack on cryptosystem is discussed in this work. 
To prevent such attacks, cyrptographic mechanisms are employed to offer security to the data 
by encrypting the data 
streams with a secret key and transform the data into a human non-readable 
format. The attempt to exercise a brute force to decrypt the information 
is exhaustive and can even be unfeasible. 
To efficiently decode the secret key and decrypt the information, 
adversaries target utilizing the information obtained through 
side-channels, termed as side-channel attacks. 
Side-channels are inherent in any given design and physical side-channel attacks exploit the information from these rather than exploiting vulnerabilities in the software. 
There exist both physical and microarchitectural side-channels that can leak secure critical information 
through acoustics, electromagnetic (EM) radiations, power trace, thermal maps and cache-access information \cite{sca_1, sca_2, sca_3, sca_4, sca_5}. 
Power signature based side-channel threats are a pivotal threat as power consumption is an inherent and preliminary characteristic of any digital system. 

This work considers a power signature based side-channel attack on encryption algorithm executing on FPGAs as they
are proliferating into data centers for compute-intensive operations such as encryption. 
For the power analysis based SCA to be successful, the attacker measures the power traces from the system while triggering crypto operations on the system. 
This trace is then studied statistically to deduce the secret key. The fundamental principle underlying this attack is that different modules (operations) of AES consume different power, 
and thereby studying the power trace reveals the operation, 
based on which the secret key can be deduced.

Pengyuan Yu et al. in \cite{Yu_codes'07} propose an intelligent place-and-route technique to facilitate symmetrical routing as a defense against power analysis SCA on FPGA. Work in \cite{Yuval_crypto'03} describes how a circuit can be 
transformed to a larger circuit to defend against probe-based physical SCAs, but, the technique proposed is very complex. Work in \cite{Kocher_crypto'99} and \cite{Chari'99} describes algorithmic countermeasures to thwart SCAs which attempts to minimize the correlation between the intermediate values and the secret key ;and by algorithmically adding noise respectively. Also, circuit-level countermeasures are presented in papers \cite{Suzuki_IEICE'07,Trichina_IACR'03,Yang_DATE'05, Dubey_host'20, Matovu_dasp'20}. It is observed that the existing defenses require modifications in physical designs, leading to larger overheads and design complexity. 

To overcome these challenges and defend against power analysis SCA, I propose 
\shrt. More details on the proposed \shrt\ are presented below. 

\subsubsection{Introduction to Correlation Power Analysis}
\label{setup}
The setup harnessed for measuring the power is described in this section followed by a brief introduction to the process of CPA (Correlation Power Analysis) analysis for key extraction.

\begin{figure}[!htb]
\figSpace
    \centering
    \includegraphics[width=1\textwidth]{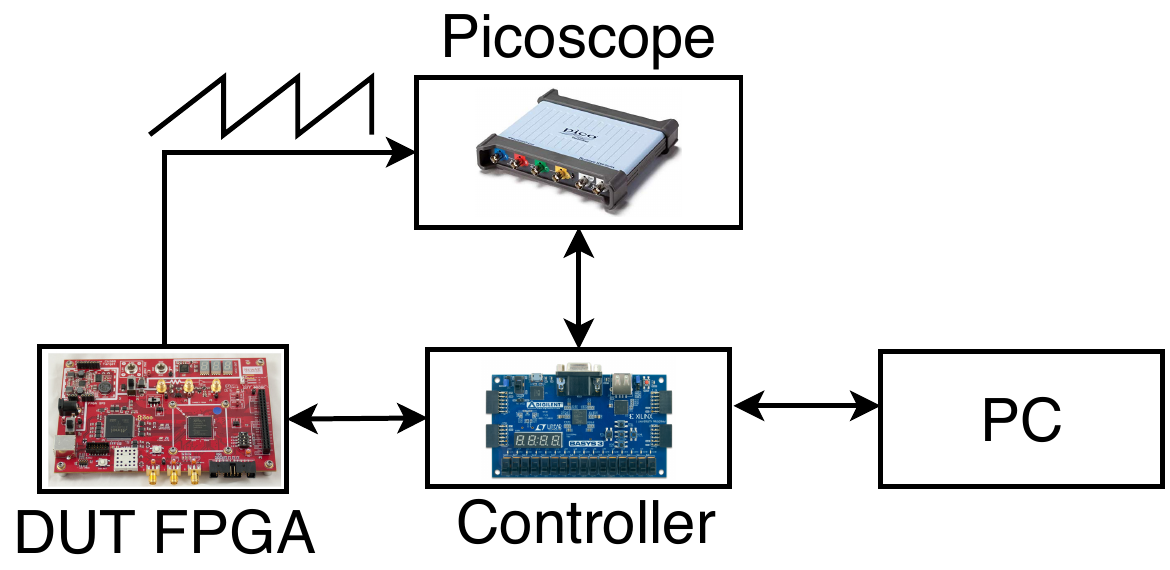}
    \caption{In-house setup for measuring the power of the DUT running AES implementation \cite{Bakry_2019, Bakry_2013}}
    \label{fig:setup}
    \figSpace
\end{figure}

\begin{figure}[!htb]
\figSpace
    \centering
    \includegraphics[width=0.6\textwidth]{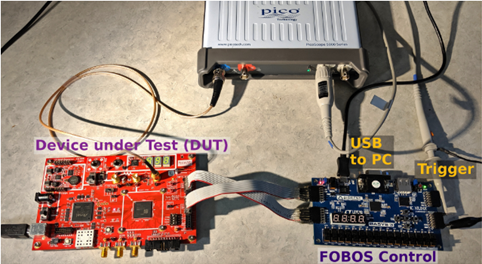}
    \caption{Actual photo of the FOBOS setup for power measurement \cite{Bakry_2019, Bakry_2013}}
    \label{fig:setup_photo}
    \figSpace
\end{figure}

\noindent \textbf{FOBOS Setup} The setup is built specially for measuring FPGA core power for physical side-channel attack analysis. The setup is termed as FOBOS (Flexible Open-source workBench fOr Side-channel analysis) \cite{Bakry_2019, Bakry_2013}. Figure \ref{fig:setup} shows the block diagram of the FOBOS setup, Figure \ref{fig:setup_photo} is a photo of the components connected together for power measurement. The Controller is an Artix-7 based FPGA that receives the test vectors from the PC and communicates with the DUT FPGA (Device Under Test) which is the target FPGA platform running the AES implemenatation. The target initiates instances of the AES cryptosystem and delivers the results to the controller. Meanwhile, the controller also triggers Picoscope measurement cycle at the same time as the AES and delivers the measured power and the cipher text to the PC. The Picoscope captures the entire trace of the AES cycle and keeps iterating for every new AES cycle. The PC runs the Python scripts that are responsible for sending the test vectors and key (secret key) to the controller and accumulating all the results in a numpy array. The FOBOS is completely reconfigurable to suit the specific needs of the measurements and application. 
The setup is described in more detail in \cite{Bakry_2019, Bakry_2013}.

\label{proposed}

\begin{figure*}[!htb]
\figSpace
    \centering
    \includegraphics[width=1\textwidth]{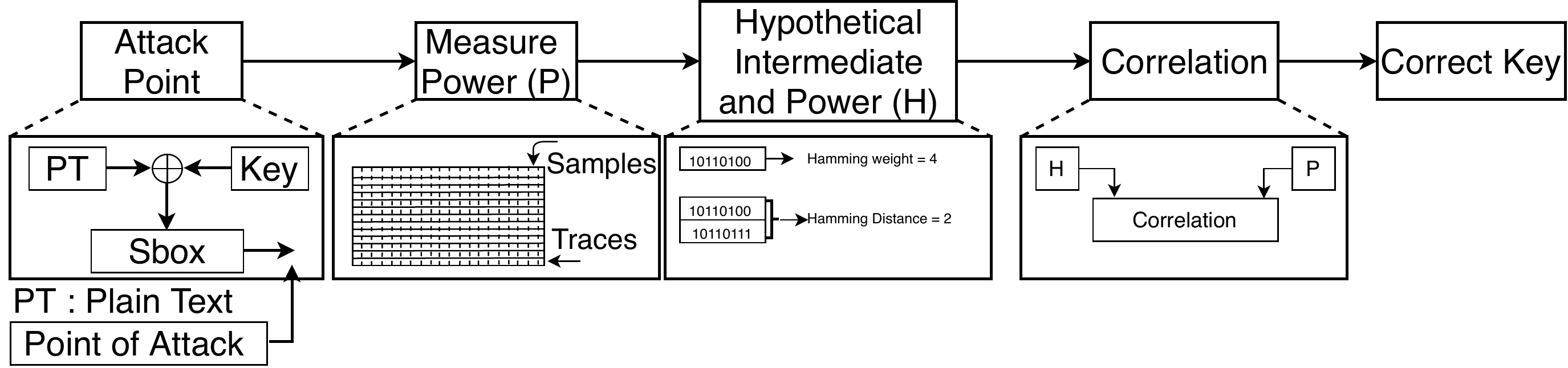}
    \caption{The process of Correlation Power Analysis (CPA) to extract the correct sequence of key used by the AES}
    \label{fig:process}
    \figSpace
\end{figure*}

\begin{figure*}[!htb]
\figSpace
    \centering
    \includegraphics[width=1\textwidth]{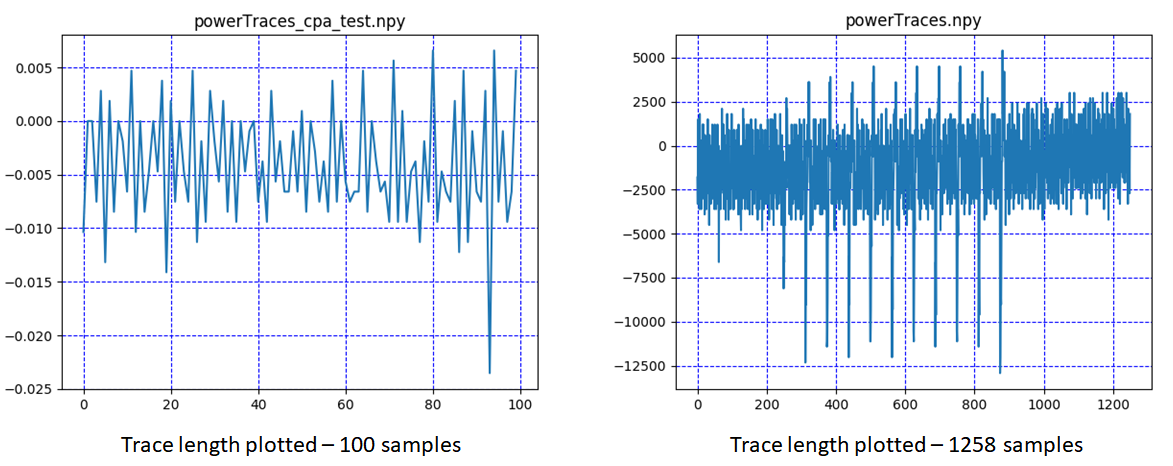}
    \caption{Measured power traces with two different sample sizes}
    \label{fig:sample_traces}
    \figSpace
\end{figure*}

\begin{figure*}[!htb]
\figSpace
    \centering
    \includegraphics[width=1\textwidth]{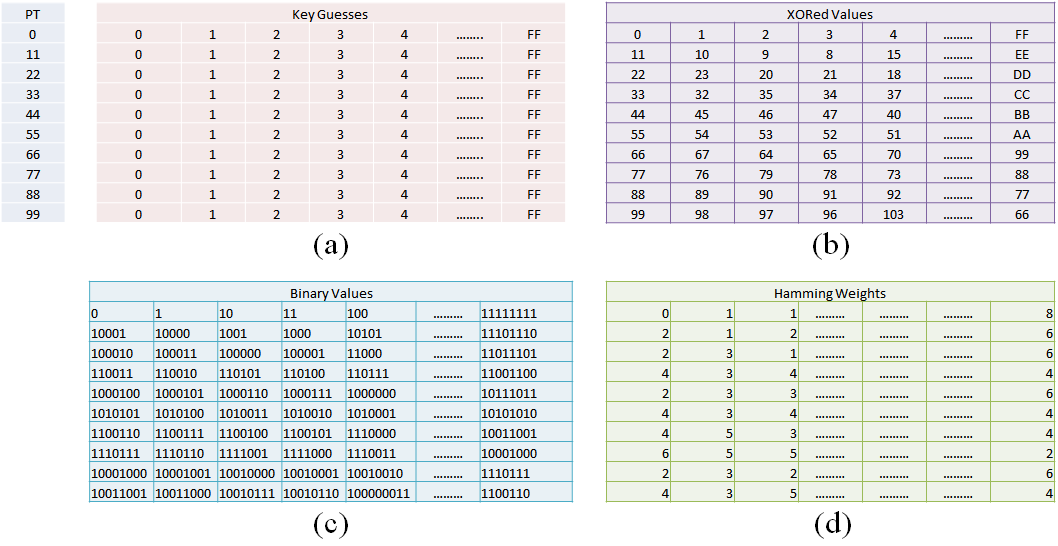}
    \caption{Correlation Power Analysis (CPA) steps}
    \label{fig:cpa_steps}
    \figSpace
\end{figure*}

\begin{figure*}[!htb]
\figSpace
    \centering
    \includegraphics[width=1\textwidth]{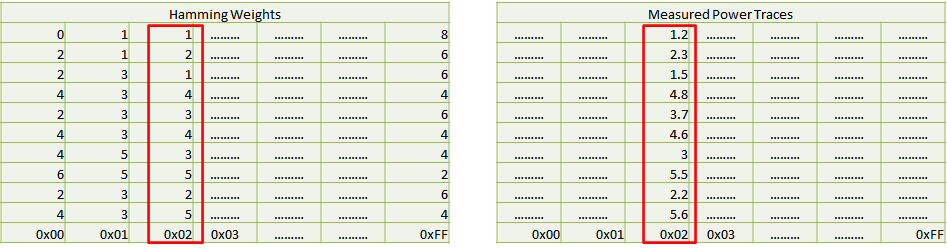}
    \caption{Example of correlation between measured and hypothetical power}
    \label{fig:correlation_matching}
    \figSpace
\end{figure*}

\noindent \textbf{Attack Model}
The attack model and the assumptions for a successful attack have been described below: 
\begin{enumerate}
    \item Physical access: The adversary needs to have physical access to the cryptosystem for obvious reasons - the CPA analysis needs power traces as one of its inputs to calculate key bytes. The physical access taps the power input to the FPGA so the traces can be captured. 
    
    \item Access to PT: The attacker also has access to the plaintext(PT) which is used by the system. 
    
    \item Access to the AES implementation: The adversary needs to have an idea of how the AES has been implemented internally. This is needed to choose the appropriate attack point and decide whether only PT and power traces are sufficient to succeed the attack phase. 
    
    \item AES timing: It is helpful for the attacker to have access to the time it takes for the intermediate values to be processed and be available at the output of the point of attack. This has been discussed in the next section. 
\end{enumerate}

\noindent \textbf{CPA Analysis} Figure \ref{fig:process} illustrates the process of how CPA analysis is performed on the system to derive the correct combination of key input. In design, the length of the plain text, cipher text and the key is 128 bits wide. The adversary must try to derive as much correct key bits as possible to break the cryptosystem security. Once the correct key has been derived, the adversary gets access to the systems where the same key was used for providing security. There are some assumptions that are precursory to the tampering of the system which have been discussed previously. The CPA attack is one of the ways in which an adversary can gain access to the AES key. Referring to Figure \ref{fig:process}, the attacker begins by deciding the point of attack. The point of attack is selected such that the value (output) available relates to the combination of the plaintext and the key (the attacker does not have access to the key, partial or whole). The block in conventional AES implementation that is chosen is the `sbox' aka substitution box. The contents of this lookup table is open sourced and hence even the attacker has access to it. As discussed previously, the attacker has access to the time it takes for the data to reach the point of attack or the sbox in this case. Proceeding further, the power to the DUT FPGA is measured and stored. Power values corresponding to one full AES cycle are known as samples, whereas, the individual runs of the AES (with different test vector, with the same key) are known as traces; sample traces with different sample sizes is shown in Figure \ref{fig:sample_traces}. The CPA then involves calculating the hypothetical intermediate and the hypothetical power, shown in Figure \ref{fig:cpa_steps}. The output of a sbox block is tried to mimic here to calculate the hypothetical power. Thing to note here is the attacker has no access to the actual key used and hence, it tries to generate all possible values (typically it is done byte wise, so a total of 256 possible values). After the output value of the sbox is known, by a combination of guessed key and plaintext, hamming weight or distance is calculated to represent power. This hypothetical power and the actual measured power are then correlated to see which power output value (hypothetical) corresponds strongly with the actual measured value and the correct key is the one that corresponds to that hypothetical value calculated previously; the process of correlation is shown in Figure \ref{fig:correlation_matching}. By iterating through this process a number of times the full key is derived.

\section{Power-Swapper: Proposed Defense Against CPA-based Side-Channel Attack}

\begin{figure*}[!htb]
    \centering
    \figSpace
    \includegraphics[width=1\textwidth]{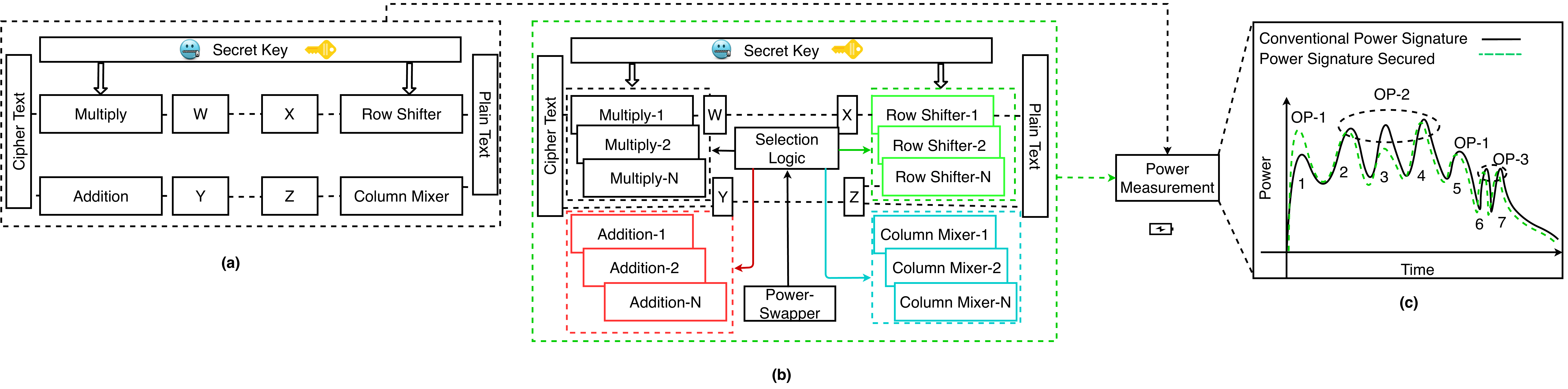}
    \caption{(a)Conventional cryptosystem where the attacker can deduce secret information from the observed power traces that correspond to the operations;(b) Cryptosystem protected by \shrt\ where the power trace does leak information but it leads to wrong information deduction by the attacker as standby modules aka approximate-modules add misleading information by leaking power traces, similar to other modules in magnitude, that are not known previously to the attacker; (c) Power signatures of the outputs of a conventional unprotected cryptosystem and a system protected by the proposed \shrt. *The figure is shown for visualization purposes and should not be used as an accurate depiction of a AES cryptosystem's internal structure*}
    \label{fig:gomac_main}
    \figSpace
\end{figure*}

The proposed \shrt\ has been outlined in Figure \ref{fig:gomac_main} where part (a) shows the internal structure of a conventional FPGA cryptosystem where each module has its own power consumption rating. The attacker then performs the power analysis on the system and then through statistical methods the adversary tries to deduce the secret key information. This is possible due to the fact that the instantaneous power consumption value would correspond to the operation of module. 
Based on these sequence of operations, the attacker can deduce the secret information. As shown in Figure \ref{fig:gomac_main}(c), the power consumption waveform has seven peaks in total each corresponding to some operation. For instance, peak 1 and 5 have the same magnitude and inferred to belong to the same operation `OP-1'; peaks 2, 3 and 4 belong to `OP-2' and so on. The information leakage in this case is maximum and it is highly correlating the
power traces which can lead to leakage of secret information.

On the contrary, Figure \ref{fig:gomac_main}(b) shows the cyrptosystem being secured by \shrt\ where the internally implemented functional modules still perform the same tasks as in a conventional FPGA shown previously except for the fact that there are other approximate-blocks that perform the same function but are designed in such a way that they have different power consumption ratings. These approximate blocks are chosen randomly during runtime by the selection logic which is controlled by the \shrt. Since each block still does the same task, there will not be any deviations in the functionality of the application. 

Power-Swapper uses physically unclonable function (PUF) block within the \shrt\ to make the selection process random and unpredictable to the attacker. 
Figure \ref{fig:gomac_main}(c) shows the waveform of the power traces corresponding to the proposed \shrt\ where some of the peaks show different values compared to the conventional FPGA power trace. Peak 1 which previously would give information of operation OP-1 now corresponds to OP-2 as per the attacker based on the power analysis. 

Peak 3, which belonged to OP-2 now corresponds to the power consumption similar to OP-1. As can be seen, the victim is not altered yet the power traces are completely different and they are not known to and which mislead the attacker. Even if the attacker tries to study a large number of patterns to find the power trace modifications injected by the approximate modules, the efforts would become futile as the trace will keep sweeping between different power magnitudes due to the randomness derived from the PUF block. The power consumption magnitudes of the alternate implementations of basic blocks range from \tx{p1} to \tx{p5} where \tx{p1 \textless\ p2\textless\ p3\textless\ p4\textless\ p5} and the range of the alternate block-1 performing the operations is in the range \tx{p1 to p3}, while that of the block-2 would be in range of \tx{p2 to p4} while yet another block would have it in \tx{p3 to p5} range. As there is overlap in the power consumption range, one approximate block's power corresponds to some other block's range when they both are swapped during runtime. 
Hence, the attacker would be forced to deduce the sequence of operations as two different ones whereas internally the same row shifting operation was performed with two different power signatures.

\begin{figure}[!htb]
\figSpace
    \centering
    \includegraphics[width=0.8\textwidth]{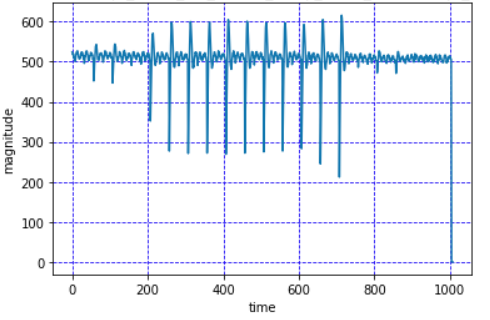}
    \caption{Power trace of AES engine implemented on FPGA}
    \label{fig:trace}
    \figSpace
\end{figure}

\begin{figure}[!htb]
\figSpace
    \centering
    \includegraphics[width=0.8\textwidth]{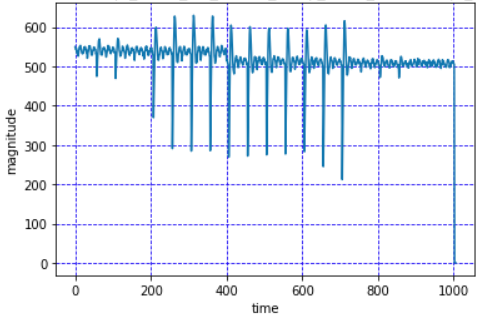}
    \caption{Power trace of AES controlled by \shrt\ . The magnitude is seen to increase(in this case) given the presence of approximate functional blocks}
    \label{fig:trace_edit}
        \figSpace
\end{figure}

Refer to Figure \ref{fig:trace} which shows the power trace of AES implementation without the proposed method. The trace was observed with the following parameters: DUT clock as 1 MHz, sampling frequency of 50 MHz and ADC sampling rate of 50MSps. If we observe the magnitude of the waveform and compare that with the magnitude shown in Figure \ref{fig:trace_edit}, the change can be vividly seen owing to the approximate modules that perform essentially the same task but with a different power consumption. The way this disrupts the CPA power analysis is: the hypothetical intermediate that will be calculated by the adversary will remain the same (as discussed previously); key and the plaintext remain the same for a particular AES cycle. On the contrary, the actual power consumption would come out to be different and hence, if not for all the parts of the key but for some, parts of the key the adversary derives is incorrect disrupting the CPA analysis. Figure \ref{fig:trace_edit} illustrates an increase in the magnitude but it can also be the opposite depending how the approximate modules are designed. Hence, theoretically, if one harnesses the approximate modules for enhancing security in FPGA based crypto implementations, power analysis based side channel attacks could be thwarted.

\subsubsection{Experimental Results}
\label{results}
The experimental setup used for capturing AES traces (without \shrt) was: 1. Artix-7 based FPGA controller, CW305 Artix FPGA Target DUT board, PicoScope 5000 series for capturing the power traces, system clock frequency used was 1 MHz. The cyrpto application implemented on the DUT board was AES \cite{Bakry_2019} with 128 bits of plaintext, key and output cipher text. Pearson correlation was used to calculate hypothetical power. Automation scripts were used to provide test vectors to the DUT and 1 million traces were collected for analysis.   

Refer to Table \ref{tbl:output} for the power traces observed by the attacker with \shrt. As can be seen from table, the information deduced by the attacker based on power consumption values are completely different compared to the actual operation executed on the core. The modified power signatures observed by the attacker are highlighted in red. The modified signatures are a result of the \shrt\ choosing one of the approximate blocks. Similarly, referring to Table \ref{tbl:result_keys}, the key derived using CPA without and with \shrt\ has been shown. The bytes/nibble that were wrongly correlated to the key guesses - as described previously - have been highlighted. These wrong portions of the keys are observed as the effect of the approximate modules introduced by \shrt\ . It is to be noted that the length of the plaintext and key does not in anyway affect the efficacy of the proposed method. The results in the Tables \ref{tbl:output}, \ref{tbl:result_keys} provide sufficient proof that by employing approximate modules, cryptosystems can be rendered resilient against power analysis based side-channel attacks.

\noindent \textbf{Ovehead Analysis}: As with every system, the proposed methodology will also have overheads. The \shrt\ requires that approximate modules be added to the original implementation of AES and these modules will be selected during the application execution. Needless to say, the modules will require additional space on the FPGA fabric along with some increase in power consumption. Switching between these modules will also lead to small overheads. The small, if not insignificant, overhead would be the trade off between security and power/area.
The resource utilization for two variants of AES is shown in Table \ref{tbl:overhead_variation_1} and \ref{tbl:overhead_variation_2}. One variant is implemented as a base version, without any pragmas in Vivado HLS 2019; Table \ref{tbl:overhead_variation_1} presents the resource utilization results. Another variant is implemented using pipeline and loop unroll pragmas; the results are presented in Table \ref{tbl:overhead_variation_2}. Both variants essentially function the same except that the power signatures are different. The total hardware resource utilization will be decided based on how many variants of AES, or blocks within AES, is utilized to include a range of power signatures for each block. The results presented in Table \ref{tbl:overhead_variation_1} and \ref{tbl:overhead_variation_2} are for reference only; they will scale based on the security to area/power tradeoff chosen as per the required resiliency against attack.

\begin{table}[!htb]
\centering
\tableSpace
\caption{Impact of \shrt\ on power trace extraction}
\label{tbl:output}
\scalebox{0.8}{
\begin{tabular}{|l||l|l|l|}
\hline
\textbf{Scenario} & \textbf{Victim Power Trace} & \multicolumn{2}{c|}{\textbf{Power Trace with \shrt\ }} \\ 
\hline
 & & \textbf{Instance-1} & \textbf{Instance-2} \\
\hline
Scenario-1 & OP-1/OP-2/OP-3/OP-4 & \textcolor{red}{OP-3/OP-2/OP-1}/OP-4 & \textcolor{red}{OP-2/OP-3/OP-4/OP-1} \\
 \hline
Scenario-2 & OP-1/OP-1/OP-4/OP-3 & \textcolor{red}{OP-2/OP-3/OP-2/OP-1} & \textcolor{red}{OP-3}/OP-1/OP-1/\textcolor{red}{OP-3} \\
\hline
\end{tabular}}
\tableSpace
\end{table}

\begin{table}[!htb]
\tableSpace
\centering
\caption{Impact of \shrt\ on keys}
\scalebox{0.8}{
\begin{tabular}{|c|c|c|}
\hline
\textbf{Trace \#} & \textbf{Correct key} & \textbf{Incorrect key with \shrt\ } \\
\hline
Trace-1 & 51720187c36e0c8523acb8535a870703 & 51\textcolor{red}{5}2\textcolor{red}{2}187c\textcolor{red}{a}6e\comm{a2}8523acb8\textcolor{red}{e}35a8707\textcolor{red}{9}3 \\ 
\hline
Trace-2 & d14a900c7391d64101fe33a85b0793cb & \comm{a1}4a90\comm{d}c7391d6\comm{32}01fe33a85b\comm{16}93cb \\
\hline
\end{tabular}}
\label{tbl:result_keys}
\tableSpace
\end{table}

\begin{table}[!htb]
\centering
\tableSpace
\caption{Resource utilization of AES without pragmas implemented on a Zedboard FPGA}
\label{tbl:overhead_variation_1}
\vspace{0.5em}
\scalebox{1}{
\begin{tabular}{|l|l|l|l|l|l|}
\hline
Name & BRAM 18K & DSP48E & FF & LUT & URAM \\
\hline
DSP & - & - & - & - & -\\
\hline
Expression & - & - & 0 & 64 & -\\
\hline
FIFO & - & - & - & - & -\\
\hline
Instance & 0 & - & 1245 & 6648 & -\\
\hline
Memory & - & - & - & - & -\\
\hline
Multiplexer & - & - & - & 183 & -\\
\hline
Register & 0 & - & 1311 & 32 & -\\
\hline
Total & 0 & 0 & 2556 & 6895 & 0\\
\hline
Available & 280 & 220 & 106400 & 53200 & 0\\
\hline
Utilization (\%) & 0 & 0 & 2 & 12 & 0\\
\hline
\end{tabular}}
\tableSpace
\end{table}

\begin{table}[!htb]
\centering
\tableSpace
\caption{Resource utilization of AES with loop unroll and pipeline pragmas implemented on a Zedboard FPGA}
\label{tbl:overhead_variation_2}
\vspace{0.5em}
\scalebox{1}{
\begin{tabular}{|l|l|l|l|l|l|}
\hline
Name & BRAM 18K & DSP48E & FF & LUT & URAM \\
\hline
DSP & - & - & - & - & -\\
\hline
Expression & - & - & 0 & 1478 & -\\
\hline
FIFO & - & - & - & - & -\\
\hline
Instance & 0 & - & 4356 & 16698 & -\\
\hline
Memory & - & - & - & - & -\\
\hline
Multiplexer & - & - & - & 162 & -\\
\hline
Register & 0 & - & 3717 & 32 & -\\
\hline
Total & 0 & 0 & 8073 & 18370 & 0\\
\hline
Available & 280 & 220 & 106400 & 53200 & 0\\
\hline
Utilization (\%) & 0 & 0 & 7 & 34 & 0\\
\hline
\end{tabular}}
\tableSpace
\end{table}

\subsubsection{Conclusion and Future Work}
\label{conclusion}
In this work, I discussed the physical power SCAs, discussed the severity of the threats posed and delineated the works in the past. In contrast to the existing works, proposed \shrt\ will preserve the victim's secret information without any modifications to the victim algorithm in itself. I hope the community will be intrigued by the preliminary results discussed in this work and I plan to develop this work in future to deliver more details of the mechanism that would benefit the security critical processes.

\section*{Acknowledgement}
I would like to thank Dr.Jens-Peter Kaps and Dr. Abubakr Abdulgadir from Cryptographic Engineering Research Group - George Mason University for providing access to the FOBOS \cite{Bakry_2019} setup for power-side channel analysis, and offering all the kind help in configuring FOBOS for the experiments. 

%% file: conclusion.tex
\section[Conclusion]{Conclusion}
This thesis discussed the importance of security to computing systems. We discussed the attacks, like reverse engineering, threatening the security community and their capabilities to disrupt design confidentiality. Further, defense against IC reverse engineering attacks was discussed which includes the automation application for logic locking defense and robustness evaluation against attacks, followed by the design of hardware communication modules. Further, physical side-channel attack on crypto systems on FPGA was discussed in detail along with a proposed defense against such attack. The thesis also includes full working code for the automation application and hardware module.